\documentclass[12pt,a4paper,twoside]{article}
  \usepackage[reqno,sumlimits]{amsmath}
  \usepackage{amsfonts,amssymb,amsthm,amsbsy,bbm,head} %macht Kopf
    \advance\textheight by \topskip
    \topmargin \paperheight \advance\topmargin by -\textheight 
    \divide\topmargin by 2 \advance\topmargin by -1in 
    \advance\topmargin by -\baselineskip
    \advance\topmargin by -\headsep
    \oddsidemargin \paperwidth \advance\oddsidemargin by -\textwidth
    \divide\oddsidemargin by 2 \advance\oddsidemargin by -1in
    \evensidemargin \oddsidemargin
\newtheorem{theorem}{Theorem}[section]
\newtheorem{proposition}[theorem]{Proposition}
\newtheorem{lemma}[theorem]{Lemma}
\newtheorem{corollary}[theorem]{Corollary}
\theoremstyle{definition}
\newtheorem{remark}[theorem]{Remark}
\newtheorem{remarks}[theorem]{Remarks}

\makeatletter
  \@addtoreset{equation}{section}

\DeclareMathOperator{\Tr}{Tr}
\def\indfkt#1{\raisebox{0.4ex}{$\chi$}_{#1}}

\DeclareMathOperator{\supp}{supp}
\DeclareMathOperator{\dist}{dist}
\DeclareMathOperator{\sgn}{sgn}
\DeclareMathOperator{\spec}{spec}

\def\esssup{\mathop{\rm ess\,sup}}
\let\Im\undefined

\DeclareMathOperator{\Im}{Im}

\def\d{{\mathrm{d}}}
\def\e{{\mathrm e}}
\def\i{{\mathrm i}}
\def\ass#1{{\rm (\sf #1\rm )}}
\newcounter{numcount}
\newcommand{\labelnummer}{\mbox{(\roman{numcount})}}%

    {\let\curlabelspeicher\@currentlabel%
     \begin{list}{\labelnummer}{\usecounter{numcount}%
                  \topsep1ex\partopsep2ex\parsep0pt\itemsep1.5ex\@plus1\p@%
                  \labelwidth3em\itemindent0em\labelsep1em%
                  \leftmargin3em}%
     \let\saveitem\item%
     \def\item{\saveitem%
               \def\@currentlabel{\curlabelspeicher\labelnummer}%
               \let\label\bemlabel}}%
   {\end{list}}%

\newenvironment{indentnummer*}%
    {\begin{list}{\labelnummer}{\usecounter{numcount}%
                  \topsep2ex\partopsep2ex\parsep0pt\itemsep1.5ex\@plus1\p@%
                  \labelwidth3em\itemindent0em\labelsep1em%
                  \leftmargin3.5em}}%
   {\end{list}}%

\newenvironment{nummer}%
    {\let\curlabelspeicher\@currentlabel%
     \begin{list}{\labelnummer}{\usecounter{numcount}\leftmargin0pt%
                  \topsep1ex\partopsep2ex\parsep0pt\itemsep1.5ex\@plus1\p@%
                  \labelwidth3em\itemindent3.5em\labelsep1em}%
     \let\saveitem\item%
     \def\item{\saveitem%
               \def\@currentlabel{\curlabelspeicher\labelnummer}%
               \let\label\bemlabel}}%
   {\end{list}}%

\def\itemref#1{\expandafter\@setref\csname r@#1item\endcsname\@firstoftwo{#1}}%

\def\bemlabel#1{\@bsphack%
  \protected@write\@auxout{}%
         {\string\newlabel{#1}{{\@currentlabel}{\thepage}}}%
  \ifmmode\else%
  \protected@write\@auxout{}%
         {\string\newlabel{#1item}{{\labelnummer}{\thepage}}}%
  \fi%
  \@esphack}%
\newif\ifper\pertrue

\def\au#1#2{#1 #2}
\def\et{ and }
\def\bti#1{\emph{#1},}
\def\ti#1{``#1'',}

\def\z{\@ifnextchar[\zz\zzz}
\def\zz[#1]#2#3#4#5{\perfalse\emph{#2} \textbf{#3} (#5) \linebreak[0]#4
    [#1].}
\def\zzz#1#2#3#4{\emph{#1} \textbf{#2} (#4) \linebreak[0]#3\ifper.\fi\pertrue}

\def\pub{\@ifstar\pubstar\pubnostar}
\def\pubnostar{\@ifnextchar[\@@pubnostar\@pubnostar}
\def\@@pubnostar[#1]#2#3#4{#3, #2, #4, #1\ifper.\fi\pertrue}
\def\@pubnostar#1#2#3{#2, #1, #3\ifper.\fi\pertrue}
\def\pubstar[#1]#2#3#4{\perfalse #3, #2,  #4 [#1].}

\makeatother
\RequirePackage{multicol} 

\makeatletter

\renewcommand\indexname{Citation Index}

\def\citationindex{\addcontentsline{toc}{section}{\indexname}%
                 \@input@{\jobname.ind}}%

\makeindex

\def\@citex[#1]#2{%
  \let\@citea\@empty
  \@cite{\@for\@citeb:=#2\do%
         {\@citea\def\@citea{,\penalty\@m\ }%
          \edef\@citeb{\expandafter\@firstofone\@citeb}%
          \if@filesw\immediate\write\@auxout{\string\citation{\@citeb}}\fi%
          \@ifundefined{b@\@citeb}%
          {\mbox{\reset@font\bfseries ?}%
             \G@refundefinedtrue\@latex@warning
             {Citation `\@citeb' on page \thepage \space undefined}}%
          {\hbox{\csname b@\@citeb\endcsname}%
            \def\speicher{\csname b@\@citeb\endcsname}%
            \makeatother%
                \index{\speicher @[\speicher]}%
            \makeatletter%
          }%  endelse 
         }% enddo
        }{#1}%  Vervollstaendigung des Aufrufs von \@cite
}%

\def\@idxitem#1,{\par\noindent#1\quad\hfill\raggedleft\hangindent 5\p@}

\makeatother

\begin{document}
%
%------------------------------------------------------------------------------
  \thispagestyle{wanda} % only for the first page
%------------------------------------------------------------------------------
\vspace*{0.5cm}
\begin{center}
{\large\bf
   EXISTENCE AND UNIQUENESS \\ OF 
   THE INTEGRATED DENSITY OF STATES \\
   FOR  SCHR{\"O}DINGER OPERATORS \\
   WITH  MAGNETIC FIELDS \\ AND UNBOUNDED RANDOM  POTENTIALS\\
}\end{center}
\vspace*{0.5cm}
\begin{center}
{\large THOMAS HUPFER {\normalsize{and}} HAJO LESCHKE}\\[3.5pt] 
{\it  Institut f\"ur Theoretische Physik, 
      Universit\"at Erlangen-N\"urnberg\\
      Staudtstr. 7, D-91058 Erlangen, Germany\\
      E-mail:
     hupfer@theorie1.physik.uni-erlangen.de\\
      leschke@theorie1.physik.uni-erlangen.de}\\
\vspace*{0.5cm}
{\large PETER M{\"U}LLER}\\[3.5pt] 
{\it  Institut f\"ur Theoretische Physik, 
      Universit\"at G\"ottingen\\
      Bunsenstr. 9, D-37073 G\"ottingen, Germany\\
      E-mail: 
      mueller@theorie.physik.uni-goettingen.de}\\
\vspace*{0.5cm}
{\large SIMONE WARZEL}\\[3.5pt] 
{\it  Institut f\"ur Theoretische Physik, 
      Universit\"at Erlangen-N\"urnberg\\
      Staudtstr. 7, D-91058 Erlangen, Germany\\
      E-mail:
      warzel@theorie1.physik.uni-erlangen.de}\\
\vspace*{0.8cm}
{Version of 23 October 2001}\\
\vskip2mm
{\emph{Dedicated to Jean-Michel Combes on the occasion of his $ 60^{\rm th} $ birthday}}\\
%\vskip2mm
%{Mathematical Physics Preprint archive: math-ph/0010013}\\
\end{center}
%\vskip4pt
%{2000 Mathematics~Subject~Classification:~35J10,~35Q40,~47B80}\\
\vspace*{0.5cm}
%
%------------------------------------------------------------------------------
\begin{center}
\begin{minipage}[t]{129mm }
\noindent
The object of the present study is the integrated density of states
of a quantum particle 
in multi-dimensional Euclidean space which is characterized
by a Schr{\"o}dinger operator with a constant magnetic field and
a random potential which may be unbounded from above and from below.
For an ergodic random potential
satisfying a simple moment condition,
we give a detailed proof that the  
infinite-volume limits of spatial eigenvalue concentrations of 
finite-volume operators with different boundary conditions exist almost surely.
Since all these limits are 
shown to coincide with the expectation of the trace of the spatially
localized spectral family of the infinite-volume operator,
the integrated density of states is almost surely non-random and independent of the chosen 
boundary condition.
Our proof of the independence of the boundary condition
builds on and generalizes certain results 
obtained by S.~Doi, A.~Iwatsuka and T.~Mine [Math.\ Z. {\bf 237} (2001) 335--371] and 
S.~Nakamura [J.\ Funct.\ Anal.\ {\bf 173} (2001) 136--152].
\end{minipage}
\end{center}
%\vskip2mm
%\noindent
%{\it Keywords and phrases:} Random magnetic Schr\"odinger operators, Integrated density of states, 
%Independence of boundary conditions. 

%------------------------------------------------------------------------------
     \newpage
     \pagestyle{myheadings} % return to enumeration of pages
     \markboth{\hspace*{1cm} \it \small T.~Hupfer, H.~Leschke, P.~M{\"u}ller \& S.~Warzel \hfill}{%
      \hfill  \it \small    Integrated Density of States for Random Schr{\"o}dinger Operators with Magnetic Fields\hspace*{1cm}}
%------------------------------------------------------------------------------
\tableofcontents
%------------------------------------------------------------------------------
\section{Introduction}
The integrated density of states is 
an important quantity in the 
theory \cite{Kir89,CaLa90,PaFi92} and 
application \cite{ShEf84,BoEn84,LiGr88,AnFoSt82,KuMeTi88}
of Schr{\"o}dinger operators 
for a particle in $d$-dimensional Euclidean space $\mathbbm{R}^d $ 
($ d = 1, 2, 3, \dots $)
subject to a random potential.
It determines the free energy of the corresponding non-interacting 
many-particle system in the thermodynamic limit 
and also enters formulas for transport coefficients.
In accordance with statistical mechanics, 
to define the integrated density of states
one usually
considers first the system confined to a bounded box.
For the corresponding finite-volume random Schr{\"o}dinger operator 
to be self-adjoint 
one then has to impose a boundary condition on the (wave) functions 
in its domain.
The infinite-volume limit of the number of eigenvalues per volume of this 
finite-volume operator below a given energy 
defines the integrated density of states~$ N $.
Basic questions are whether this limit exists, is independent of almost 
all realizations of the random potential and of the chosen boundary condition. 
These are the questions of existence, non-randomness, and uniqueness.
For vanishing magnetic field these questions were settled several years 
ago \cite{Pas71,Nak77,KirMar82,Kir89,CaLa90,PaFi92}, see also \cite{KosSch00} for
a more recent approach.
For non-zero magnetic fields the existence and non-randomness of $ N $ 
are known since \cite{Mat93,Uek94,BrHuLe93}.
Uniqueness, that is, the independence of the boundary condition 
follows  from recent results in \cite{DoIwMi01} and \cite{Nak00}
for bounded  below or bounded random potentials, respectively.
However, a proof of uniqueness
is lacking for random potentials which are unbounded from below. 

The main goal of the present paper is to give a detailed proof of the existence, non-randomness, 
and uniqueness of $ N $
for the case of constant magnetic fields and a wide class of
ergodic random potentials which may be unbounded from above as well as from below and which satisfy
a simple moment condition.
In particular, $ N $ is shown to coincide with the expectation of the trace of the spatially localized
spectral family of the infinite-volume operator.
As a consequence, the set of growth points of $ N $ is immediately identified
with the almost-sure spectrum of this operator.
Important examples of random potentials which may yield operators 
unbounded from below 
and to which our main 
result, Theorem~\ref{Thm:IDOS}, applies, are alloy-type, Poissonian, and 
Gaussian random potentials.

Our proof of the existence, non-randomness, and uniqueness of $ N $ 
differs from those outlined in \cite{Mat93,Uek94,BrHuLe93}
and is patterned on the one of analogous statements 
for vanishing magnetic fields in the monograph of Pastur and Figotin \cite{PaFi92}. 
Since the infinite-volume operator 
may be unbounded from below, 
we have to make sure that the sequence of the underlying finite-volume 
density-of-states measures is ``tight near minus infinity''. 
Our proof of the independence of the boundary condition uses an approximation argument which
reduces the problem to that of bounded random potentials and therefore
heavily relies on results of Doi, Iwatsuka and Mine \cite{DoIwMi01} or Nakamura \cite{Nak00}.
%------------------------------------------------------------------------------
%------------------------------------------------------------------------------
%
\section{Random Schr{\"o}dinger Operators with Constant Magnetic Fields}
\subsection{Basic Notation}
As usual, let $\mathbbm{N} := \{1,2,3,\ldots\}$ denote the set of natural
numbers. 
Let $\mathbbm{R}$, respectively $\mathbbm{C}$,
denote the algebraic field of real, respectively complex, numbers. An
open cube $\Lambda$ in $d$-dimensional 
Euclidean space $\mathbbm{R}^{d}$, $ d \in \mathbbm{N}$, is a translate of
the $d$-fold Cartesian product $ I\times\cdots\times I$ of
an open interval $I\subseteq\mathbbm{R}$. 
The open unit cube in $ \mathbbm{R}^d $ which is
centered at site $y = (y_1 , \dots , y_d ) \in \mathbbm{R}^d$ and whose edges are oriented parallel
to the co-ordinate axes is the product $ \Lambda(y) := {\mathsf X}_{j=1}^d ] y_j - 1/2, y_j + 1/2 [$
of open intervals.
We call a bounded open cube $ \Lambda $ \emph{compatible} with the 
(structure of the simple cubic) lattice 
$\mathbbm{Z}^d$
if it is the interior of the closure of a union of finitely many open unit cubes
centered at lattice sites, that is
\begin{equation}\label{Def:compat}
  \Lambda = \Bigl(\; \overline{ 
      \bigcup_{y\in \Lambda \cap \mathbbm{Z}^d} \Lambda(y)} \;\Bigr)^{\rm int}.
\end{equation}
The Euclidean norm of
$x\in\mathbbm{R}^{d}$ is denoted by 
$ |x|:=\big(\sum_{j=1}^{d}x_{j}^2\big)^{1/2} $.
We denote the volume of a Borel subset $\Lambda\subseteq\mathbbm{R}^{d}$
with respect to the $d$-dimensional Lebesgue measure as $|\Lambda| :=
\int_{\Lambda}\!\d^{d}x = \int_{\mathbbm{R}^d}\!\d^{d}x \, \indfkt{\Lambda}(x)$
where $ \indfkt{\Lambda} $ is the indicator function of $ \Lambda $.
In particular, if $ \Lambda $ is the strictly positive half-line,
$ \Theta := \indfkt{]\, 0 , \infty[} $ is the left-continuous Heaviside 
unit-step function.  
We use the notation $ \alpha \mapsto ( \alpha -1 )! $ for
     Euler's gamma function \cite{GrRy}.
The Banach space ${\rm L}^{p}(\Lambda)$ consists
of the Borel-measurable complex-valued functions 
$f: \Lambda\to\mathbbm{C}$ which are identified if their values differ only on a
set of Lebesgue measure zero and possess a finite ${\rm L}^p$-norm 
\begin{equation}
  \left| f \right|_p := \left\{
    \begin{array}{ll} \displaystyle
      \biggl(\int_{\Lambda}\!\d^{d}x\; |f(x)|^{p}\biggr)^{1/p}\;
      & \text{if}~p \in [1, \infty[\,,\\[2.5ex]
      \displaystyle\esssup\limits_{x\in\Lambda} |f(x)|
      & \text{if}~p=\infty\,.
    \end{array}
  \right.
\end{equation}
We recall that ${\rm L}^{2}(\Lambda)$ is a
separable Hilbert space with scalar product   
$  \langle f, g \rangle := \int_{\Lambda}\d^{d}x\,
  \overline{f(x)} $\hspace{0pt}$\, g(x)
$.
The overbar denotes complex conjugation. 
We write $f \in
{\rm L}^{p}_{\rm loc}(\mathbbm{R}^{d})$, if $ f \in {\rm L}^{p}(\Lambda)$ 
for any bounded Borel set $\Lambda\subset \mathbbm{R}^{d}$.
Moreover,  
$\mathcal{C}^{n}_{0}(\Lambda) $  stands for the vector space of
functions $f: \Lambda\to\mathbbm{C} $ 
which are $n$ times continuously  
differentiable and have compact supports.
The vector
space of functions which have compact supports and are continuous, respectively 
arbitrarily often differentiable, is denoted by 
$\mathcal{C}_{0}(\Lambda)$, respectively $\mathcal{C}^{\infty}_{0}(\Lambda)$. 
Finally, $ W^{1,2}(\Lambda) :=  \big\{ \phi \in {\rm L}^2(\Lambda) : 
      \nabla \, \phi \in ({\rm L}^2(\Lambda))^d \big\}$  
is the first-order Sobolev space of $ {\rm L}^2 $-type where $ \nabla $ stands for the 
gradient in the sense of distributions on $\mathcal{C}^{\infty}_{0}(\Lambda)$. 

The absolute value of a closed operator 
$ F: \mathcal{D}(F) \to {\rm L}^2(\Lambda)$, 
densely defined with domain $ \mathcal{D}(F) \subseteq {\rm L}^2(\Lambda) $
and adjoint $ F^\dagger $,
is the positive operator 
$ \left| F \right| := ( F^\dagger F )^{1/2} $.
The (uniform) norm of a bounded operator $ F : {\rm L}^2(\Lambda) \to {\rm L}^2(\Lambda)$ is 
defined as $ \left\| F \right\| := \sup\big\{ \left| F f \right|_2 \, : \, 
  f \in {\rm L}^2(\Lambda)\, ,  \left| f \right|_2 =1 \big\} $.  
Finally, for $  p \in [1 , \infty[ $ 
we will use the notation
\begin{equation}\label{eq:notation}
  \left\| F \right\|_p := \Big( \Tr \left| F \right|^{p} \Big)^{1/p}
\end{equation}
for the  
(von Neumann-) Schatten norm of an 
operator $ F $ on $ {\rm L}^2(\Lambda) $ in the Banach space 
$ \mathcal{J}_p\big({\rm L}^2(\Lambda)\big) $.
For these $ \mathcal{J}_p $-spaces of compact operators, 
see \cite{Sim79TR,BirSol87}.
In particular, $  \mathcal{J}_1 $ is the space of trace-class and $  \mathcal{J}_2 $ the space of 
Hilbert-Schmidt operators.
%
%%%%%%%%%%%%%%%%%%%%%%%%%%%%%%%%%%%%%%%%%%%%%%%%%%%%%%%%%%%%%%%%%%%%%%%%%%%%
%
%
\subsection{Basic Assumptions and Definitions of the Operators}
Let $ (\Omega, \mathcal{A}, \mathbbm{P}) $ be a complete probability space
and $ \mathbbm{E}\{ \cdot \} := \int_{\Omega} \! \mathbbm{P}(\d\omega)( \cdot ) $ be the expectation
induced by the probability measure $\mathbbm{P}$. 
By a \emph{random potential} we mean a (scalar) random field
$ V:\, \Omega \times \mathbbm{R}^{d} \to \mathbbm{R} \, $,
$ (\omega ,x)\mapsto V^{(\omega )}(x) $ 
which is assumed to be jointly measurable with respect to the product of the 
sigma-algebra $ \mathcal{A} $ of event sets in $\Omega$ 
and the sigma-algebra $\mathcal{B}(\mathbbm{R}^d)$ 
of Borel sets in $ \mathbbm{R}^d $. 
We will always assume $ d \geq 2 $, because magnetic fields in one space 
dimension may be ``gauged away'' and are therefore of no physical relevance.
Furthermore, for $ d = 1 $ far more is known \cite{CaLa90,PaFi92} thanks to
methods which only work for one dimension.\\

\noindent
We list three properties which $ V $ may have or not:
\begin{indentnummer*}
\item[\ass{S}] There exists some pair of reals 
  $ p_1 > p(d) $ and $ p_2 > p_1 d / \left[2 ( p_1 - p(d) ) \right]$ 
  such that 
  \begin{equation}\label{Eq:I}
    \sup_{y \in \mathbb{Z}^d} \, \mathbbm{E} \Big\{ \big[
    \int_{\Lambda(y)} \!\!\! \d^d x \, |V(x)|^{\, p_1} \big]^{p_2/p_1} \Big\} 
    < \infty.
  \end{equation}
  Here $ p(d) $ is defined as follows:
  $p(d) := 2$ if $d \leq 3 $, $ p(d) := d/2$ if $d \geq 5$ and $ p(4) > 2$, otherwise arbitrary.
  
\item[\ass{E}] $ V $ is $ \mathbbm{Z}^d $-ergodic or $ \mathbbm{R}^d $-ergodic.
\item[\ass{I}] 
  $ V $ satisfies the finiteness condition
  \begin{equation}\label{eq:2.5}
    \sup_{y \in \mathbb{Z}^d} \, \mathbbm{E} \big[
    \int_{\Lambda(y)} \!\!\!  \d^d x \, |V(x)|^{2\vartheta+1} \big] < \infty,
  \end{equation}
  where $\vartheta \in \mathbbm{N}$ is the smallest integer with $ \vartheta > d/4$.
\end{indentnummer*}
\begin{remarks}
  \begin{nummer}
  \item
    Property \ass{E} requires the existence of a group 
    $ \mathcal{T}_x $, $x \in \mathbbm{Z}^d$  or 
    $  \mathbbm{R}^d $,
    of probability-preserving and ergodic transformations
    on $ \Omega $ such that $ V $ is 
    $ \mathbbm{Z}^d $- or $ \mathbbm{R}^d $-homogeneous in the sense that
    $ V^{(\mathcal{T}_x \omega)}(y) = V^{(\omega)}(y-x) $ for all $ x \in \mathbbm{Z}^d $ or $ \mathbbm{R}^d $, 
    all $ y \in \mathbbm{R}^d $, and all $ \omega \in \Omega $; see \cite{Kir89}. 
  \item\label{Rem:Omegas}
    Property \ass{S} assures that the realization $V^{(\omega)}: x \mapsto V^{(\omega)}(x) $ of $ V $ belongs to 
    ${\rm L}^{p(d)}_{\rm loc}(\mathbbm{R}^d)$ for each $ \omega $ in some subset $\Omega_{\sf S} \in \mathcal{A}$ 
    of $ \Omega $ with full probability, in symbols, 
    $\mathbb{P}(\Omega_{\sf S}) = 1$.
    If $ d \neq 4 $, property~\ass{I} in general does
    not imply property~\ass{S} even if property~\ass{E} is supposed.
    Given \ass{E}, a sufficient criterion for both \ass{S} and \ass{I} to hold 
    is the finiteness
    \begin{equation}\label{Eq:I+D}
      \mathbbm{E} \big[
      \int_{\Lambda(0)} \!\!\!  \d^d x \, |V(x)|^{p} \, \big] < \infty
    \end{equation}
    for some real $ p > d +1 $.
    To prove this claim for property \ass{S} 
    we choose $p_1 = p_2 = p $ in (\ref{Eq:I}).
    For \ass{I} the claim follows from $ 2\vartheta \leq d $. 
    If the random potential is $ \mathbbm{R}^d $-homogeneous, 
    Fubini's theorem gives $
    \mathbbm{E} \big[|V(0)|^{\, p} \big] $ for the l.h.s.\ of (\ref{Eq:I+D}).
  \end{nummer}
\end{remarks}
In the present paper we mainly consider 
the case of a \emph{constant magnetic field} in $ \mathbbm{R}^d $. This is characterized
by a skew-symmetric tensor 
with real constant components $B_{jk} $, $j, k \in \{1, \dots , d\}$.
On account of gauge equivalence, there is no loss of generality in 
assuming that the  \emph{vector potential} $A : \mathbbm{R}^d \to  \mathbbm{R}^d$, $x \mapsto A(x)$,
generating the magnetic field 
according to $ B_{jk} = \partial_j A_k - \partial_k A_j $, satisfies property 
\begin{indentnummer*}
\item[\ass{C}] 
  $A $ is the vector potential of a constant magnetic field 
  in the symmetric gauge, that is, 
  its components are given by  
  $ A_k(x) =  \frac{1}{2} \sum_{j=1}^d x_j B_{jk}$ with $ k \in \{1, \dots , d\}$. 
\end{indentnummer*}
%
%\subsection{Definition of the Operators}
%
We are now prepared to precisely define 
magnetic Schr{\"o}dinger operators with random
potentials on the Hilbert spaces $ {\rm L}^2(\Lambda)$ and $ {\rm L}^2(\mathbbm{R}^d) $. 
The finite-volume case is treated in 
\begin{proposition}\label{Prop:DefH}
  Let $\Lambda \subset \mathbbm{R}^d$ be a bounded 
    open cube. Let $ A $ be a vector potential with property \ass{C} and $V$ be a random 
  potential with property \ass{S}. {\rm [}Recall from Remark~\ref{Rem:Omegas} the definition of the set\/ $ \Omega_{\sf S} $.{\rm ]}   
  Then
  \begin{indentnummer*}
  \item
    the sesquilinear form
    $
      \mathcal{Q} \times \mathcal{Q} \ni ( \varphi , \psi ) \mapsto 
     \frac{1}{2} \sum_{j=1}^{d} 
      \left\langle ( {\i} \nabla_j + A_j) \, \varphi \, , 
        \, ( {\i} \nabla_j + A_j) \, \psi \right\rangle
    $
    with form domain $\mathcal{Q}= W^{1,2}(\Lambda) $ or
     $\mathcal{Q}=  \mathcal{C}^\infty_0(\Lambda) $ is positive, symmetric and closed, respectively closable. 
    Accordingly, both forms uniquely define positive self-adjoint operators on ${\rm L}^2(\Lambda)$ 
    which we denote by $ H_{\Lambda,{\rm N}}(A,0)$ and $ H_{\Lambda,{\rm D}}(A,0)$, respectively. 
  \item
    the two operators 
    \begin{equation}
      H_{\Lambda,{\rm X}}(A,V^{(\omega)}) := H_{\Lambda,{\rm X}}(A,0) + V^{(\omega)}, 
      \qquad {\rm X=D}\;\; {\rm or} \;\; {\rm X=N}, 
    \end{equation}
    are well defined  on ${\rm L}^2(\Lambda)$ as form sums
    for all $\omega \in \Omega_{\sf S}$, hence for $ \mathbbm{P} $-almost all $ \omega \in \Omega $.
    They are self-adjoint and bounded below. 
    Moreover, the mapping 
    $H_{\Lambda,X}(A, V): \Omega_{\sf S}  \ni \omega \mapsto H_{\Lambda,{\rm X}}(A,V^{(\omega)})$
    is measurable.
    We call it the 
    \emph{finite-volume magnetic Schr{\"o}dinger operator with random potential $V$} 
    and Dirichlet or Neumann boundary condition if $\rm X = D$ or 
    $\rm X = N$, respectively.
  \item
    the spectrum of $H_{\Lambda,\rm X}(A,V^{(\omega)})$ is purely discrete for all 
    $ \omega \in \Omega_{\sf S} $ such that
    the (random)  
    \emph{finite-volume density-of-states measure}, defined by
    \begin{equation}\label{Def:nu}
      \nu_{\Lambda,{\rm X}}^{(\omega)}(I):=
      \Tr\Big[ \indfkt{I}\big(H_{\Lambda,{\rm X}}(A ,V^{(\omega)})\big) \Big],
    \end{equation}
    is a positive Borel measure on the real line $\mathbbm{R}$ for all $\omega \, \in \, \Omega_{\sf S}$.  
    Here,  $\indfkt{I}\big(H_{\Lambda,{\rm X}}(A ,V^{(\omega)})\big)$ is 
    the spectral projection operator of $H_{\Lambda,{\rm X}}(A ,V^{(\omega)})$ 
    associated with the energy regime $I\in \mathcal{B}(\mathbbm{R})$. 
    Moreover, the
    (unbounded left-continuous) distribution function
    \begin{equation}\label{eq:finite-volumeIDOS}
      N_{\Lambda , \rm X}^{(\omega)}(E) :=  
      \nu_{\Lambda,{\rm X}}^{(\omega)}\big(\left] - \infty, E \right[ \big) 
      = \Tr\Big[ \Theta\big( E - H_{\Lambda,{\rm X}}(A ,V^{(\omega)}) \big) \Big] < \infty
    \end{equation}
    of $ \nu_{\Lambda,{\rm X}}^{(\omega)} $,
    called the \emph{finite-volume integrated density of states},
    is finite for all energies $ E \in \mathbbm{R} $. 
\end{indentnummer*}
\end{proposition}
\begin{proof}
  The assumptions of Proposition~\ref{Prop:DefH} imply those of \cite[Prop.~2.1]{HuLeMuWa01}. 
\end{proof}
\begin{remark}
  Counting multiplicity, $\nu_{\Lambda,{\rm X}}^{(\omega)}(I)$ is just the number of 
  eigenvalues of the operator $H_{\Lambda,\rm X}(A ,V^{(\omega)})$ 
  in the Borel set $ I \subseteq \mathbbm{R} $. Since this number is almost surely finite for every
  bounded $ I $, the mapping 
  $ \nu_{\Lambda,{\rm X}} : \Omega_{\sf S} \ni \omega \mapsto  \nu^{(\omega)}_{\Lambda,{\rm X}} $ 
  is a random Borel measure in the sense that $ \nu^{(\omega)}_{\Lambda,{\rm X}} $ 
  assigns a finite length to each bounded 
  Borel set.
\end{remark}
The infinite-volume case is treated in
\begin{proposition}\label{Prop:DefH1}
  Let $ A $ be a vector potential with property \ass{C} and $V$ be a random 
  potential with property \ass{S}. Then
  \begin{indentnummer*}
    \item
      the 
      operator $ \mathcal{C}_0^\infty(\mathbbm{R}^d) \ni \psi \mapsto 
        \frac{1}{2} \sum_{j = 1}^d ({\i} \partial_j + A_j)^2 \, \psi + V^{(\omega)}\psi
        $
      is essentially self-adjoint
      for all $\omega \in \Omega_{\sf S} $. Its self-adjoint closure 
      on ${\rm L}^2(\mathbbm{R}^d)$
      is denoted by $H(A,V^{(\omega)})$. 
     \item
       the mapping
       $H(A, V): \Omega_{\sf S}  \ni \omega \mapsto H(A,V^{(\omega)})$
       is measurable. 
       We call it the 
       \emph{infinite-volume magnetic Schr{\"o}dinger operator with random potential $V$}.
     \end{indentnummer*}
   \end{proposition}
\begin{proof}
  See for example \cite[Prop.~2.2]{HuLeMuWa01}.
\end{proof}
%%%%%%%%%%%%%%%%%%%%%%%%%%%%%%%%%%%%%%%%%%%%%%%%%%%%%%%%%%%%%%%%%%%%%%%%
%
\begin{remarks}
  \begin{nummer}
    \item
      For alternative or weaker criteria instead of \ass{S} guaranteeing 
      the almost-sure self-adjointness
      of $ H(0, V) $, see \cite[Thm.~5.8]{PaFi92} or \cite[Thm.~1 on p.~299]{Kir89}.
    \item
      The infinite-volume magnetic 
      Schr{\"o}dinger operator without scalar potential, $H(A,0)$, 
      is unitarily invariant under so-called \emph{magnetic translations} \cite{Zak64,Lev70}.
      The latter form a family of unitary operators 
      $\{T_x\}_{x \in \mathbbm{R}^d}$ on ${\rm L}^2(\mathbbm{R}^d)$
      defined by
      \begin{equation}
        \left(T_x \psi\right) (y) := \exp\Bigg[\frac{\i}{2} \sum_{j,k=1}^d (x_j - y_j) B_{jk} \, x_k  \Bigg]\, \psi(y - x ) , 
        \qquad \qquad \psi \in {\rm L}^2(\mathbbm{R}^d).
      \end{equation}
      In the situation of Proposition~\ref{Prop:DefH1}
      and if the random potential $V$ has property \ass{E}, we have
      \begin{equation}\label{eq:magntrans}
        T_x \, H(A, V^{(\omega)}) \, T_x^\dagger =  H(A, V^{(\mathcal{T}_x \omega)})
      \end{equation}
      for all $ \omega \in \Omega_{\sf S}$ and all $ x \in \mathbbm{Z}^d $ or $ x \in \mathbbm{R}^d $, 
      depending on whether $ V $
      is $ \mathbbm{Z}^d $- or $ \mathbbm{R}^d $-ergodic. Hence, following standard arguments, $ H(A, V) $ is an
      \emph{ergodic operator} and its spectral components are non-random,
      see \cite[Thm. 2.1]{Uek94}. Moreover, the discrete spectrum of 
      $ H(A, V^{(\omega)}) $ is empty for $\mathbbm{P}$-almost 
      all $\omega\in \Omega$, 
      see \cite{Kir89,CaLa90,Uek94}, 
      because the family $ \{T_x\}_{x \in \mathbbm{Z}^d} $ and hence
      $  \{T_x\}_{x \in \mathbbm{R}^d} $ is \emph{total}. 
      The latter is true by definition, since the 
      subset $ \{T_x \psi  \}_{ x \in  \mathbbm{Z}^d } \subset {\rm L}^2(\mathbbm{R}^d)$ 
      contains an infinite set 
      of pairwise orthogonal functions 
      for each $ \psi \in \mathcal{C}_0(\mathbbm{R}^d) $ which is dense in $ {\rm L}^2(\mathbbm{R}^d) $.
    \end{nummer}
\end{remarks}
%
%%%%%%%%%%%%%%%%%%%%%%%%%%%%%%%%%%%%%%%%%%%%%%%%%%%%%%%%%%%%%%%%%%%%%%%%%%%%
%%%%%%%%%%%%%%%%%%%%%%%%%%%%%%%%%%%%%%%%%%%%%%%%%%%%%%%%%%%%%%%%%%%%%%%%%%%%
%
\section{The Integrated Density of States}
\subsection{Existence and Uniqueness} 
The quantity of main interest in the present paper is 
the integrated density of states
and its corresponding measure, called the density-of-states measure.
The next theorem deals with its definition and its 
representation as an infinite-volume 
limit of the suitably scaled finite-volume counterparts~(\ref{eq:finite-volumeIDOS}).
It is the main result of the present paper. 
\begin{theorem}\label{Thm:IDOS}
  Let $\Gamma \subset \mathbbm{R}^d $ be a bounded open cube 
  compatible with the lattice $ \mathbbm{Z}^d $ {\rm [}recall (\ref{Def:compat}){\rm ]}
  and let
  $\indfkt{\Gamma}$ denote the multiplication operator
  associated with the indicator function of $ \Gamma $.
  Assume that the potentials $A$ and $V$ have the
  properties \ass{C}, \ass{S}, \ass{I}, and \ass{E}.
  Then the \emph{(infinite-volume) integrated density of states} 
  \begin{equation}\label{Eq:DefIDOS}
    N(E) := \frac{1}{\left|\Gamma\right|} \, \mathbbm{E}\Big\{
    \Tr\big[ \indfkt{\Gamma} \, \Theta\big(E - H(A,V) \big) 
    \, \indfkt{\Gamma} \big]\Big\} < \infty
  \end{equation}
  is well defined for all energies $ E \in \mathbbm{R} $ 
  in terms of the spatially localized spectral family of 
  the infinite-volume operator $ H(A, V) $.
  It is 
  the (unbounded left-continuous) distribution function 
  of some positive Borel measure %$ \nu $ 
  on the real line $ \mathbbm{R}$ 
  and independent of $ \Gamma $.
  Moreover, 
  let $\Lambda \subset \mathbbm{R}^d$ stand for bounded open cubes 
  centered at the origin.  
  Then there is a set $ \Omega_0 \in \mathcal{A} $ 
  of full probability, $ \mathbbm{P}(\Omega_0) = 1 $, such that 
  \begin{equation}\label{Eq:IDOSconv}
    N(E) =  \lim_{\Lambda \uparrow \mathbb{R}^d} \, 
    \frac{N_{\Lambda,{\rm X}}^{(\omega)}(E)}{\left| \Lambda \right|}
  \end{equation}
  holds for  both  boundary conditions ${\rm X} = {\rm D}$
  and ${\rm X} = {\rm N}$, all $ \omega \in \Omega_0 $, and 
  all $ E \in \mathbbm{R} $ 
  except the (at most countably many) discontinuity points of $ N $.
\end{theorem}
\begin{proof}
  See Section~\ref{AppIDOS}.  
\end{proof}
\begin{remarks}
  \begin{nummer}
  \item As to the limit $ \Lambda \uparrow \mathbb{R}^d $, 
    we here and in the following think of a sequence of open 
    cubes centered at the origin whose edge lengths tend to infinity. 
    But there exist
    more general sequences of expanding regions in $ \mathbbm{R}^d $ 
    for which the theorem
    remains true, see for example \cite[Rem.~1 on p.~105]{PaFi92} and \cite[p.~304]{CaLa90}.
  \item\label{Rem:Rdergod} 
    The homogeneity of the random potential and the
    magnetic field with respect to $\mathbbm{Z}^d$ 
    renders the r.h.s.\ of (\ref{Eq:DefIDOS})
    independent of $\Gamma$.
    In case $V$ is even $\mathbbm{R}^d$-ergodic, one may pick an arbitrarily
    shaped bounded subset $\Gamma \in \mathcal{B}(\mathbbm{R}^d) $ with 
    $\left| \Gamma \right| > 0$
    or even any non-zero square-integrable function instead of the 
    indicator function; for details see the next corollary.
  \item\label{Re:IDOS} 
    A proof of the \emph{existence} of the infinite-volume limits in (\ref{Eq:IDOSconv})
    under slightly different hypotheses was
    outlined in \cite{Mat93}. It uses functional-analytic arguments
    first presented in \cite{KirMar82} for the case $ A = 0 $.
    A different approach to the existence of these limits for $ A \neq 0 $,
    using Feynman-Kac(-It{\^o}) functional-integral representations of
    Schr{\"o}dinger semigroups \cite{Sim79,BrHuLe00}, 
    can be found in \cite{Uek94,BrHuLe93}.
    It dates back to \cite{Pas71,Nak77} for the case $ A = 0 $ and,
    to our knowledge, works straightforwardly in the case $ A \neq 0 $ 
    for $ \rm X = \rm D $ only. 
    For $ A \neq 0 $ \emph{uniqueness} of the infinite-volume limit in
    (\ref{Eq:IDOSconv}), that is, its independence of the boundary
    condition $ \rm X $ (previously claimed without proof in \cite{Mat93}) follows 
    from \cite{Nak00} if  
    the random potential $ V $ is bounded and from \cite{DoIwMi01} if $ V $ is
    bounded from below. 
    So the main new point about Proposition~\ref{Thm:IDOS} is that it establishes existence and uniqueness for 
    a wide class of $ V $ unbounded from below. This class also includes many $ V $ yielding operators $ H(A,V) $
    which are unbounded from below.
    Even for $ A = 0 $, Proposition~\ref{Thm:IDOS} is partially new 
    in that the corresponding result \cite[Thm.~5.20]{PaFi92}, only shows vague convergence of the underlying measures, 
    see Lemma~\ref{Thm:DOSmass} and Remarks~\ref{Rem:Pastur0} below.
  \item
    Property~\ass{S} is only assumed to guarantee the almost-sure 
    essential self-adjoint\-ness
    of the infinite-volume operator on $ \mathcal{C}_0^\infty(\mathbbm{R}^d) $.
    Property~\ass{I} is mainly technical. It ensures the existence of a sufficiently high 
    integer moment of $ V $
    needed for the applicability of standard resolvent techniques. 
    In particular, \ass{I} does not distinguish between 
    the positive part $ V_+ := \max \{ 0, V \} $ and the negative
    part $ V_- := \max \{ 0, - V \} $ of $ V $. 
    This stands in contrast to proofs based on functional-integral representations, 
    which require much stronger assumptions on $ V_- $ 
    but much weaker assumptions on $ V_+ $, see \cite[Thm.~3.1]{Uek94}.
    Instead of constant magnetic fields as demanded by property~\ass{C}, 
    the subsequent proof in Sect.~\ref{AppIDOS} can be extended straightforwardly 
    to cover also ergodic random
    magnetic fields as in \cite{Mat93,Uek94}.
  \item
    The convergence (\ref{Eq:IDOSconv}) holds for any other boundary condition $ \rm X $
    for which the self-adjoint operator $ \, H_{\Lambda, \rm X}(A,V^{(\omega)})\,  $
    obeys the inequalities $ \, H_{\Lambda, \rm N}(A,V^{(\omega)}) \; \leq \;  
    H_{\Lambda, \rm X}(A,V^{(\omega)}) \leq
    H_{\Lambda, \rm D}(A,V^{(\omega)}) $ in the sense of forms \cite[Def.~on p.~269]{ReSi78}. 
    This follows from the min-max principle \cite[Sec.~XIII.1]{ReSi78} which implies that
    the finite-volume integrated density of states $N_{\Lambda, \rm X}^{(\omega)}$ associated with  
    $ H_{\Lambda, \rm X}(A,V^{(\omega)}) $
    obeys the sandwiching estimates 
    \begin{equation}
      0 \leq N_{\Lambda, \rm D}^{(\omega)}(E) \leq
      N_{\Lambda, \rm X}^{(\omega)}(E) \leq
      N_{\Lambda, \rm N}^{(\omega)}(E) < \infty
    \end{equation}
    for every bounded open cube 
    $\Lambda \subset \mathbbm{R}^d$ and all energies  
    $E \in \mathbbm{R}$. 
  \item
    Under the assumptions of Theorem~\ref{Thm:IDOS} 
    there is some $ \Omega_1 \in \mathcal{A} $ with 
    $ \mathbbm{P}(\Omega_1) = 1 $  such that
    \begin{equation}\label{eq:alternative}
      N(E) = \lim_{\Lambda \uparrow \mathbbm{R}^d} \, \frac{1}{|\Lambda |} \,
      \Tr \big[ \indfkt{\Lambda} \, \Theta\big(E - H\big(A,V^{(\omega)}\big)\big) \, 
      \indfkt{\Lambda} \big] 
    \end{equation}
    for all $ \omega \in \Omega_1 $ and all $ E \in \mathbbm{R} $ 
    except the (at most countably many) discontinuity points of $ N $.
    This follows from the fact that 
    $  \Tr \big[ \indfkt{\Lambda} \, \Theta\big(E - H\big(A,V^{(\omega)}\big)\big) \, 
    \indfkt{\Lambda} \big] = \sum_{j \in \Lambda \cap \mathbbm{Z}^d}  
    \Tr \big[ \indfkt{\Lambda(j)} \, \Theta\big(E - H\big(A,V^{(\omega)}\big)\big) \, 
    \indfkt{\Lambda(j)} \big] $ 
    for all bounded open cubes  $ \Lambda $
    which are compatible with the lattice $  \mathbbm{Z}^d $, 
    the Birkhoff-Khintchine ergodic theorem 
    in the formulation of \cite[Prop.~1.13]{PaFi92}
    and the considerations in \cite[p.~80]{Kir85a}.
    Alternative representations of the integrated density of states 
    as in (\ref{eq:alternative}) 
    seem to date back to \cite{AvrSim83}, 
    see also \cite{CyFr87,Kir89,CaLa90,Wan95}.
  \item
    As a by-product, 
    our proof of Theorem~\ref{Thm:IDOS} yields 
    (see (\ref{eq:IDOSEndl2}) below) 
    the following rough upper
    bound on the low-energy fall-off of $N$,
    \begin{equation}\label{eq:Lifshits}
      N(E) \leq C \, \left| E \right|^{d/2 -2\vartheta}
    \end{equation}
     for all $ E \in ] - \infty, -1] $ with some constant $ C \geq 0 $, 
     see also \cite[Thm.~5.29]{PaFi92} 
     for the case $ A = 0 $. 
     The true leading behavior of $ N(E) $ for $ E \to - \infty $ is, of course, consistent with (\ref{eq:Lifshits}), but 
     typically much faster.
     For example, 
     in the case of a Gaussian random potential, in the sense of Subsection~\ref{Subsec:Ex} below, 
     it is known that $ \lim_{E \to - \infty} E^{-2} \log N(E) = - (2 C(0) )^{-1} $, also in the presence of a 
     constant magnetic field \cite{Mat93,BrHuLe93,Uek94}.
     The leading low-energy behavior 
     is less universal in case of a positive Poissonian potential 
     and a constant magnetic field \cite{BrHuLe95,Erd98,HuLeWa99,HuLeWa00,Erd00,War01},
     where $ N $ vanishes for negative energies anyway. 
     In this context we recall from \cite{Mat93,Uek94} 
     that the high-energy asymptotics is neither affected 
     by the magnetic field nor by the random potential and given by
     $\lim_{E \to \infty} E^{-d/2} \, N(E) = [ (d/2) ! \, ( 2 \pi )^{d/2} ]^{-1} $
     in accordance with Weyl's celebrated asymptotics for the free particle [59].%\cite{Wey12}. 
   \item
     In case $ H(A,V) $ is unbounded from below almost surely and serves as the one-particle Hamiltonian of a 
     macroscopic system of  non-interacting (spinless) fermions, the corresponding free energy 
     and resulting basic thermostatic quantities may nevertheless be well defined, provided that $ N(E) $ falls off to zero
     sufficiently fast as $ E \to - \infty $.
     An at least algebraic decay in the sense that $ N(E) \leq C \, \left| E \right|^{d/2 -2\alpha} $ with sufficiently large 
     $ \alpha \in \mathbbm{N} $, $ \alpha > \vartheta $, is assured by simply requiring the ergodic random potential $ V $ 
     to satisfy (\ref{eq:2.5}) 
     with $ \vartheta $ replaced by $ \alpha $. The proof of this assertion follows the same lines of reasoning 
     leading to (\ref{eq:Lifshits}).
  \end{nummer}
\end{remarks}
%
%%%%%%%%%%%%%%%%%%%%%%%%%%%%%%%%%%%%%%%%%%%%%%%%%%%%%%%%%%%%%%%%%%%%%%%%%%%%
%
%
In analogy to \cite[Prob.~II.4]{PaFi92} Theorem~\ref{Thm:IDOS} implies
\begin{corollary}\label{Cor:fFormel}
  Assume that the potentials $A$ and $V$ have the
  properties \ass{C}, \ass{S}, and \ass{I}. Moreover, let $ V $ be 
  $ \mathbbm{R}^d $-ergodic
  (and not only $ \mathbbm{Z}^d $-ergodic).
  Then
  \begin{equation}\label{eq:fformel}
    N(E) = \frac{1}{\left|f \right|_2^2} \, \mathbbm{E}\Big\{
    \Tr\Big[ \overline{f} \, \Theta\big(E - H(A,V) \big) 
      \, f \Big]\Big\}, \qquad E \in \mathbbm{R},
  \end{equation}
  for any non-zero $ f \in { \rm L}^2(\mathbbm{R}^d) $ 
  which is to be understood as a multiplication operator inside the trace.
\end{corollary}
%%%%%%%%%%%%%%%%%%%%%%%%%%%%%%%%%%%%%%%%%%%%%%%%%%%%%%%%%
\begin{remark}
  Assume the situation of Corollary~\ref{Cor:fFormel} and that the spectral projection $ \Theta( E - H(A,V)) $
  possesses $ \mathbbm{P} $-almost surely a jointly continuous integral kernel 
  $ \mathbbm{R}^d \times \mathbbm{R}^d \ni (x,y) \mapsto \Theta( E - H(A,V))(x,y) \in \mathbbm{C} $.
  Then (\ref{eq:fformel}) with $ f \in \mathcal{C}_0(\mathbbm{R}^d) $ gives by \cite[Lemma on pp.~65--66]{ReSi79}, 
  Fubini's theorem,  
  and the $ \mathbbm{R}^d $-homogeneity of $ V $ the formula
  \begin{equation}\label{Eq:intkern}
    N(E) = \mathbbm{E}\Big[ \Theta\big(E - H(A,V) \big)(0,0) \Big], \qquad E \in \mathbbm{R},
  \end{equation}
  see also \cite[Prop.~3.2]{Uek94}.
  A sufficient condition for the existence and continuity of the integral kernel \cite[Remark~6.1.(ii)]{BrHuLe00} is 
  that $ V_- $ and $ V_+ \indfkt{\Lambda} $ belong  for any bounded
  $ \Lambda \in \mathcal{B}(\mathbbm{R}^d) $
  $ \mathbbm{P} $-almost surely to the Kato class 
  \begin{equation}
    \mathcal{K}(\mathbbm{R}^d ) := \Big\{ v : \mathbbm{R}^d \to \mathbbm{R} \, : \;\,  
                     v \;\, \mbox{Borel measurable and} \;\, \lim_{t \downarrow 0 } \varkappa_t(v) = 0 \; \Big\},
  \end{equation}
  where $ \varkappa_t(v) := \sup_{x \in \mathbbm{R}^d} 
  \int_0^t \! \d s \, \int_{\mathbbm{R}^d} \! \d^d \xi \, \e^{- | \xi |^2 }   | v(x +  \xi \sqrt{s}) | $.
  While property~\ass{S} implies $ V_+ \indfkt{\Lambda} \in \mathcal{K}(\mathbbm{R}^d ) $, 
  it does not ensure
  $ V_- \in \mathcal{K}(\mathbbm{R}^d ) $ even when combined with property~\ass{I}.
  This is in agreement with the fact that $ H(A, V) $ would else be bounded from below, which, for example,
  is not the case if $ V $ is a Gaussian random potential (in the sense of Subsection~\ref{Subsec:Ex} below). 
  For weaker conditions which ensure the validity of (\ref{Eq:intkern}) for rather general random potentials including
  Gaussian ones, see \cite{BrMuLe01}.
 \end{remark} 
\begin{proof}[\bf Proof of Corollary~\ref{Cor:fFormel}]
  We may assume $ f \geq 0 $, because the general 
  case  $ f \in { \rm L}^2(\mathbbm{R}^d) $ follows therefrom.
  Let $ \left( f_n \right)_{n \in \mathbbm{N}} \subset {\rm L}^2(\mathbbm{R}^d) $ be a monotone increasing
  sequence,
  $f_n \leq f_m $ if $ n \leq m $,
  of positive simple functions approximating $ f $.
  More precisely, these functions are assumed to be of the form
  $ f_n(x) = \sum_{k =1}^n f_{n,k} \indfkt{\Gamma_{n,k}}(x) $ with suitable
  constants $  f_{n,k} \geq 0 $ and bounded Borel sets 
  $ \Gamma_{n,k} \in \mathcal{B}(\mathbbm{R}^d) $ 
  which are pairwise disjoint for each fixed $ n $.
  Using (\ref{Eq:DefIDOS}) and the  
  $ \mathbbm{R}^d $-homogeneity of the random potential 
  (see Remark~\ref{Rem:Rdergod}) one verifies that (\ref{eq:fformel})
  is valid for all simple functions.
  Thanks to the convergence $ f_n \to f $ as $ n \to \infty $ 
  in $ {\rm L}^2(\mathbbm{R}^d) $
  this implies
  \begin{equation}\label{eq:simplef}
    \lim_{n,m \to \infty} \, \int_{\Omega} \! \mathbbm{P}(\d \omega) \,\,
    \big\| \Theta^{(\omega)}
    \big(f_n - f_m \big) \big\|_2^2 =
    N(E) \, \lim_{n,m \to \infty} \, \big| f_n - f_m \big|_2^2 = 0,
  \end{equation}
  where we are using the abbreviation $\Theta^{(\omega)} := \Theta\big(E - H(A,V^{(\omega)}) \big) $.
  Hence there exists some  sequence $ \left( n_j \right)_{j \in \mathbbm{N}} $
  of natural numbers such that 
  \begin{equation}\label{eq:3.10}
    \mathbbm{E} \Big[ \| \Theta \,
    f_{n_{j+1}} - \Theta \, f_{n_j}  \|_2 \Big] 
    \leq \Big\{ \mathbbm{E} \Big[ \| \Theta \,
    f_{n_{j+1}} - \Theta \, f_{n_j}  \|_2^2 \Big] \Big\}^{1/2} \leq 2^{-j} 
  \end{equation}
  for all $ j \in \mathbbm{N} $ by Jensen's inequality and (\ref{eq:simplef}).
  Thanks to monotonicity the r.h.s. of the estimate
  \begin{equation}\label{eq:fabschaetz}
     \|\Theta^{(\omega)}
     f_{n_{i}} - \Theta^{(\omega)} f_{n_j}  \|_2 
     \leq \sum_{k= \min\{i , j \}}^\infty  \| \Theta^{(\omega)}
     f_{n_{k+1}} - \Theta^{(\omega)} f_{n_{k}} \|_2,
   \end{equation}
   converges pointwise for all $ \omega \in \Omega_{\sf S} $ as $ i, j \to \infty $.
   Since $ \lim_{i,j \to \infty} \sum_{k= \min\{i , j \}}^\infty 
  \mathbbm{E}\big[ \| \Theta\,
  f_{n_{k+1}} $\hspace{0pt}$- \Theta\, f_{n_{k}}  \|_2 \big] = 0 $ by (\ref{eq:3.10}), the 
  monotone- and dominated-convergence theorems
  imply that the r.h.s.\ (and hence the l.h.s.) of (\ref{eq:fabschaetz}) 
  converges in fact to zero 
  for $ \mathbbm{P} $-almost all $ \omega \in \Omega $. 
  In other words, the subsequence
  $\big( \Theta^{(\omega)} f_{n_j} \big)_j $ is
  Cauchy in $ \mathcal{J}_2({\rm L}^2(\mathbbm{R}^d)) $
  for $ \mathbbm{P} $-almost all $ \omega \in \Omega $.
  Since the space $ \mathcal{J}_2({\rm L}^2(\mathbbm{R}^d)) $ is complete, 
  this sequence converges with respect to
  the Hilbert-Schmidt norm $ \left\| \cdot \right\|_2 $ to some 
  $ F^{(\omega)} \in \mathcal{J}_2({\rm L}^2(\mathbbm{R}^d)) $.
  Let $ f : \mathcal{C}_0^\infty(\mathbbm{R}^d) \to {\rm L}^2(\mathbbm{R}^d) $ 
  denote a multiplication operator associated with $ f $.
  The above convergence and 
  $ \lim_{n \to \infty } \, \left| (f - f_n) \psi \right|_2 = 0 $
  for all $ \psi \in \mathcal{C}_0^\infty(\mathbbm{R}^d) $ imply that
  $ F^{(\omega)} $ is the unique continuous extension
  of $ \Theta^{(\omega)} f $ from 
  $ \mathcal{C}_0^\infty(\mathbbm{R}^d) $ to the whole Hilbert space 
  $ {\rm L}^2(\mathbbm{R}^d) $. Denoting this extension also by 
  $  \Theta^{(\omega)} f $, we thus have
  \begin{equation}\label{eq:HSconverg}
    \lim_{j \to \infty } \, \big\| \Theta^{(\omega)} 
    f_{n_j} -  \Theta^{(\omega)}  f \big\|_2 = 0
  \end{equation}
  for $ \mathbbm{P} $-almost all $ \omega \in \Omega $. 
  We therefore get 
  \begin{multline}
    \mathbbm{E} \Big[ \big\|  \Theta\big(E - H(A,V) \big) f \big\|_2^2 \Big]
    =
    \mathbbm{E} \Big[ \lim_{j \to \infty} \, 
    \big\|  \Theta\big(E - H(A,V) \big) f_{n_j} \big\|_2^2 \Big] \\
    =  \lim_{j \to \infty} \,\mathbbm{E} \Big[  
    \big\|  \Theta\big(E - H(A,V) \big) f_{n_j} \big\|_2^2 \Big]
    = N(E) \, \lim_{j \to \infty } \, \big| f_{n_j} \big|_2^2 
    = N(E) \,  \left| f \right|_2^2. 
  \end{multline}
  For the second equality we used the monotone-convergence theorem.
  Note that $ (\| \Theta^{(\omega)} f_n \|_2^2 )_n $ is monotone 
  increasing 
  since $ \| \Theta^{(\omega)} f_n \|^2_2 =\| \Theta^{(\omega)} f_n^2 \Theta^{(\omega)} \|_1$.
\end{proof}
% 
%%%%%%%%%%%%%%%%%%%%%%%%%%%%%%%%%%%%%%%%%%%%%%%%%%%%%%%%
\subsection{Some Properties of the Density-of-States Measure}
The proof of Theorem~\ref{Thm:IDOS}
will be based on the (almost-sure) vague convergence \cite[Def.~30.1]{Bau92} of 
the two \emph{spatial eigenvalue concentrations}
$ \left| \Lambda \right|^{-1} \, \nu_{\Lambda, \rm X}^{(\omega)} $, with $ \rm X = \rm D $ or $ \rm X = \rm N $,
to the same non-random
measure $ \nu $ in the infinite-volume limit $ \Lambda \uparrow \mathbbm{R}^d $.
This measure is called the density-of-states measure and 
uniquely corresponds to the integrated density of states 
(\ref{Eq:DefIDOS}) in the sense that
$ N(E) = \nu ( ] - \infty , E [) $ for all $ E \in \mathbbm{R} $.
\begin{lemma}\label{Thm:DOSmass}
  Assume the situation of Theorem~\ref{Thm:IDOS}.
  Then 
  the \emph{(infinite-volume) density-of-states measure}
   \begin{equation}\label{EqDos=Dos}
    \nu(I) := \frac{1}{\left|\Gamma\right|} \, \mathbbm{E}\Big\{
    \Tr\big[ \indfkt{\Gamma} \, \indfkt{I}(H(A,V)) 
      \, \indfkt{\Gamma} \big]\Big\}, \qquad  I \in  \mathcal{B}(\mathbbm{R}),
  \end{equation} 
  is a positive Borel measure on the 
  real line $\mathbbm{R} $, 
  well defined in terms of the spatially localized projection-valued spectral measure
  of the infinite-volume random Schr{\"o}dinger operator, and independent 
  of $ \Gamma $. Moreover, in the sense of vague convergence
  \begin{equation}\label{Eq:DOSvag}
    \nu = \lim_{\Lambda \uparrow \mathbb{R}^d} \, 
    \frac{ \nu^{(\omega)}_{\Lambda,{\rm X}}}{\left| \Lambda \right|}
  \end{equation}
  for  both ${\rm X} = {\rm D}$
  and ${\rm X} = {\rm N}$ and $ \mathbbm{P} $-almost all $ \omega \in \Omega $.
\end{lemma}
\begin{proof}
  See Section~\ref{AppIDOS}.
\end{proof}
\begin{remarks}\label{Rem:Pastur0}
  \begin{nummer}
  \item\label{Rem:Pastur}
    Lemma~\ref{Thm:DOSmass} generalizes 
    \cite[Thm.~5.20]{PaFi92} which deals with the case $A=0$. In fact,
    our proof in Section~\ref{AppIDOS} closely follows 
    the arguments given there. Concerning the independence of $ \rm X $ of the
    infinite-volume limit in (\ref{Eq:DOSvag}), 
    we build on a result in \cite{Nak00}
    for bounded $ V $. (Alternatively, one may use a result in  \cite{DoIwMi01}.)
  \item 
    Lemma~\ref{Thm:DOSmass} alone does not imply the existence 
    of the integrated density of states $ N $. 
    Moreover, even if the finiteness of 
    $ \nu(] - \infty, E [) $ for all $ E \in \mathbbm{R} $
    were known, see (\ref{eq:Lifshits}), the vague convergence 
    (\ref{Eq:DOSvag}) alone would not imply 
    the pointwise convergence (\ref{Eq:IDOSconv})
    of the distribution functions in case their supports are not uniformly bounded from below.
    The latter occurs for random potentials with realizations $ V^{(\omega)} $ which yield operators $ H(A, V^{(\omega)}) $ 
    unbounded from below.
    On the other hand, (\ref{Eq:IDOSconv}) implies (\ref{Eq:DOSvag}), see 
    Proposition~\ref{Prop:Kurt} below.
  \end{nummer}
\end{remarks}

%
%%%%%%%%%%%%%%%%%%%%%%%%%%%%%%%%%%%%%%%%%%%%%%%%%%%%%%%%
%
Using (\ref{EqDos=Dos}) one may relate properties of the 
density-of-states measure $ \nu $
to simple spectral properties of the infinite-volume 
magnetic Schr{\"o}dinger operator. 
Examples are the support of $\nu$ and the 
location of the almost-sure spectrum of $ H(A, V^{(\omega)}) $ or  
the absence of a point component in the Lebesgue decomposition of $\nu $ and 
the absence of ``immobile eigenvalues'' of $H(A, V^{(\omega)}) $.
This is the
content of
\begin{corollary}\label{Thm:Stet}
  Under the assumptions of Theorem~\ref{Thm:IDOS} 
  and letting $ I \in \mathcal{B}(\mathbbm{R}) $
  the following
  equivalence holds:
  ~$  \nu(I) = 0 $~ if and only if ~$\indfkt{I}\big(H(A, V^{(\omega)})\big) = 0 $~
   for $ \mathbb{P}$-almost all $\omega\in \Omega$.
  This immediately implies:
  \begin{indentnummer*}
    \item
      ~$ \supp \nu = \spec H(A, V^{(\omega)}) $ for $\mathbb{P}$-almost all $\omega \in \Omega$.
      {\rm [}Here $ \spec H(A, V^{(\omega)}) $ denotes the  spectrum of  $  H(A, V^{(\omega)}) $ and 
      ~$\supp \nu := \{ E \in \mathbbm{R} :  \nu(]E - \varepsilon, E + \varepsilon[) > 0 \;\;
      \mbox{\rm for all} \; \varepsilon > 0 \} $ is 
      the topological support of $ \nu $.{\rm ]} 
    \item
      ~$0 = \nu(\{E\}) \; \big( = \lim_{\varepsilon \downarrow 0} \left[ N( E + \varepsilon) - N(E) \right] \big)$ 
      ~if and
      only if $E\in \mathbbm{R}$ is not an eigenvalue of $H(A,V^{(\omega)})$
      for $\mathbb{P}$-almost all $\omega \in \Omega$.
  \end{indentnummer*}
\end{corollary}
\begin{proof}
  If $ \indfkt{I}\big(H(A, V^{(\omega)})\big) = 0 $ for $ \mathbb{P}$-almost all 
  $\omega \in \Omega $, then
  $ \nu(I) = 0 $ using
  (\ref{EqDos=Dos}). Conversely, 
  for every $\psi \in \mathcal{C}_0(\mathbbm{R}^d) \subset {\rm L}^2(\mathbbm{R}^d)$, 
  normalized in the sense 
  $\langle \psi , \psi \rangle = 1$, there exists a bounded open cube
  $\Gamma \subset \mathbbm{R}^d$ compatible with $ \mathbbm{Z}^d $ such that $\supp \psi \subseteq
  \Gamma$ and therefore
  \begin{equation} 
    \left\langle \psi \, , \, \indfkt{I}\big(H(A,V^{(\omega)})\big) \, \psi \right\rangle 
    \leq \Tr \left[ \indfkt{\Gamma} \, \indfkt{I}\big(H(A,V^{(\omega)})\big) \, 
      \indfkt{\Gamma} \right].
  \end{equation} 
  Taking the probabilistic expectation on both sides and using
  (\ref{EqDos=Dos}) we arrive at the sandwiching estimate $0 \leq \mathbbm{E} \left[ \langle
    \psi \, , \, \indfkt{I}(H(A,V)) \, \psi \rangle \right] \leq
  |\Gamma| \, \nu(I) = 0$ by the assumption $\nu(I)=0$.  Since
  the magnetic translations $\{T_x\}$ with
  $ x \in \mathbbm{R}^d $ or $ \mathbbm{Z}^d $ are total, the proof of \cite[Lemma~V.2.1]{CaLa90}
  shows that
  $\indfkt{I}(H(A,V^{(\omega)})) = 0$ for $\mathbbm{P}$-almost all
  $\omega\in \Omega$. 
\end{proof}
\begin{remark}
  The equivalence~(ii) of the above corollary is a continuum analogue 
  of \cite[Prop.~1.1]{CrSi83}, see also \cite[Thm.~3.3]{PaFi92}.  In the
  one-dimensional case \cite{Pas80} and the multi-dimensional lattice 
  case \cite{DelSou84}, the equivalence has been exploited to show in case $ A = 0 $ the
  (global) continuity of the integrated density of states $ N $ 
  under practically no further assumptions on the random potential beyond %\attention
  those ensuring the existence of $ N $. The proof of such a statement in the
  multi-dimensional continuum case is considered an important open
  problem \cite{Sim00}. 
  In case $ A \neq 0 $ one certainly needs additional assumptions
  as \cite{DoMa99} illustrates.
  Under certain additional assumptions
  the integrated density of states
  is not only continuous but even (locally) H\"older continuous of arbitrary order 
  strictly smaller than one \cite{CoHiNa01,HiKl01}
  or even equal to one \cite{CoHi96,BaCoHi97a,BaCoHi97b,HuLeMuWa01}.   
  The latter is equivalent to $ N $ being absolutely continuous with 
  locally bounded derivative \cite[Chap.~7, Exc.~10]{Rud87}.
\end{remark}
\subsection{Examples}\label{Subsec:Ex}
In this subsection we list three examples of (possibly unbounded) random potentials
to which the
results of the preceding subsections can be applied.
While the first one models (crystalline) disordered alloys, the other two model
(non-crystalline) amorphous solids. 
These are typical examples considered in the literature. 
Each of them is characterized by one of the following properties.
We recall from properties~\ass{S} and \ass{I} the definitions of 
the constants $ p(d) $ and $ \vartheta  $.%
\begin{indentnummer*}
  \item[\ass{A}]
    $ V $ is an \emph{alloy-type random field}, that is, 
    a random field with realizations given by
    \begin{equation}
      V^{(\omega)}(x) = \sum_{j \in \mathbbm{Z}^d} \lambda_j^{(\omega)}
      u(x -j). 
    \end{equation}
    The random variables $\{ \lambda_j \} $ are $ \mathbbm{P} $-independent 
    and identically distributed according to the common
    probability measure $ \mathcal{B}(\mathbbm{R}) \ni I \mapsto \mathbbm{P}\{ \lambda_0 \in I \} $.
    Moreover, we suppose that the Borel-measurable function
    $ u: \mathbbm{R}^d \to \mathbbm{R}  $ satisfies the Birman-Solomyak condition
    ~$ \sum_{j \in \mathbbm{Z}^d} \big(\int_{\Lambda(j)} \! \d^d x \, |u(x)|^{p_1} \big)^{1/p_1} < \infty $
    with some real $ p_1 \geq 2 \vartheta + 1 $ and that 
    $ \mathbbm{E}\left( | \lambda_0 | ^{p_2} \right) < \infty $ for some real $ p_2 $ 
    satisfying $ p_2  \geq 2 \vartheta + 1 $ and $ p_2 > p_1 d /[2 (p_1 -p(d) )] $. 
  \item[\ass{P}]
    $ V $ is a \emph{Poissonian field}, that is, a random field with realizations given by
    \begin{equation}
       V^{(\omega)}(x) = \int_{\mathbbm{R}^d} \!   \mu_{\varrho}^{(\omega)}(\d^d y) \, u( x - y),
    \end{equation}
    where $ \mu_{\varrho} $ denotes the (random) Poissonian measure 
    on $ \mathbbm{R}^d $ with parameter $ \varrho \geq 0 $. 
    Moreover, we suppose that the Borel-measurable function $ u: \mathbbm{R}^d \to \mathbbm{R} $ 
    satisfies the Birman-Solomyak condition  
    $  \sum_{j \in \mathbbm{Z}^d} 
    \big(\int_{\Lambda(j)} \! \d^d x \, |u(x )|^{2\vartheta+1} \big)^{1/(2\vartheta+1)} < \infty $.
    %where $ \vartheta \in \mathbbm{N} $ is the smallest integer with $ \vartheta > d/4 $.
  \item[\ass{G}]
    $ V $ is a \emph{Gaussian random field} \cite{Adl81,Lif95}
    which is $ \mathbbm{R}^d $-homogeneous.
    It has  zero mean, $ \mathbbm{E}\left[\,V(0)\right] = 0 $, 
    and its covariance function
    $ x \mapsto C(x):= \mathbbm{E}\left[\, V(x) V(0) \right] $ 
    is continuous at the origin where it obeys 
    $ 0 < C(0) < \infty $.
\end{indentnummer*}
The following remarks further explain the above three examples.
\begin{remarks}
  \begin{nummer}
  \item
    Consider an \emph{alloy-type random potential}, that is, a random potential with 
    property~\ass{A}. Such a potential models a (generalized) disordered alloy \cite[Ch.~21]{Kit96} 
    which is composed of different atoms occupying, at random, the sites 
    of the lattice $ \mathbbm{Z}^d \subset \mathbbm{R}^d $.
    Which kind of atom at site $ j \in \mathbbm{Z}^d $ actually interacts with the quantum particle 
    (classically) located at $ x \in \mathbbm{R}^d $
    through the potential $ \lambda_j^{(\omega)} u(x- j) $, is determined by the value $ \lambda_j^{(\omega)} \in \mathbbm{R} $ 
    of the coupling
    strength at site $ j $. 
    An alloy-type random potential $ V $ is $ \mathbbm{Z}^d $-ergodic and hence has property~\ass{E}.
    Moreover, $ V $ is a random field of the form (\ref{eq:Vallg}) below, since one may choose 
    $ \mu $ there as the random signed pure-point measure given by 
    $ \mu^{(\omega)} = \sum_{j \in \mathbbm{Z}^d } \lambda_j^{(\omega)} \, \delta_j  $
    where $ \delta_y $ denotes the Dirac measure on $ \mathbbm{R}^d $ supported at $ y \in \mathbbm{R}^d $. 
    Lemma~\ref{lem:mom} below with $ q = p_1 $ and $ r = p_2 $ shows that 
    $ V $ has property~\ass{S}. Choosing $ q = r = 2 \vartheta + 1 $ it is seen to obey property~\ass{I}. 
    %In particular, Lemma~\ref{lem:mom}  
%    yields that the assumptions on $ u $ and the distribution of $ \lambda_0 $
%    may be optimized when considering specific alloy-type random potentials.
  \item
    Consider a \emph{Poissonian potential}, that is, a random potential with 
    property~\ass{P}. Then $ V $ is $ \mathbbm{R}^d $-ergodic and hence has property~\ass{E}.
    Using the fact that the Poissonian measure $ \mu_{\varrho} $ is a 
    random Borel measure which is pure point and positive-integer valued, each realization of $ V $ is informally given by 
    $  V^{(\omega)}(x) = \sum_j u\big( x - x_j^{(\omega)} \big) $. 
    Here the Poissonian points $\{ x_j^{(\omega)} \} $
    are interpreted as the positions of impurities, each of them generating the same potential $ u $.
    The random variable $ \mu_{\varrho}(\Lambda) $ then equals the number of 
    impurities in the bounded Borel set $ \Lambda \subset \mathbbm{R}^d $ and is distributed according
    to Poisson's law
    \begin{equation} 
      \mathbbm{P}\big( \, \mu_{\varrho}(\Lambda) = n \, \big) 
      = \frac{\left( \varrho \left| \Lambda \right| \right)^{n}}{n!} \, 
      \e^{ - \varrho \left| \Lambda \right|}, \qquad  n \in \mathbbm{N} \cup \{ 0 \},
    \end{equation}
    so that
    the parameter $ \varrho $ is identified as the mean spatial concentration of impurities.
    By choosing $ \mu = \mu_{\varrho} $, $ q = p_1 = 2 \vartheta + 1 $, and $ r = p_2 > p_1 d /[ 2 p_1 - p(d) ] $
    in Lemma~\ref{lem:mom} below, one verifies
    that a Poissonian potential statisfies property~\ass{S}. Moreover, choosing $ q = r =  2 \vartheta + 1 $
    there, it is seen to obey property~\ass{I}. If $ u \geq 0 $, (\ref{Eq:intkern}) holds for the Poissonian potential.
 \item
    Consider a random field with the Gaussian property~\ass{G}. 
    Then its covariance function $ C $ is bounded and 
    uniformly continuous on $ \mathbbm{R}^d $. 
    Consequently, \cite[Thm.\ 3.2.2]{Fer75} implies the existence of 
    a separable version $ V $ of this field which is jointly  measurable.
    Speaking about a \emph{Gaussian random potential},
    it is tacitly assumed that only this version will be dealt with.
    By the Bochner-Khintchine theorem \cite[Thm.~IX.9]{ReSi80} there is a one-to-one correspondence
    between Gaussian random potentials and finite positive (and even)
    Borel measures on $ \mathbbm{R}^d $. 
    Using the identity 
    \begin{equation}
      \mathbbm{E} \left[ \left| V(0) \right|^{ p} \right] 
      = \frac{1}{[2 \pi C(0) ]^{1/2}} \int_{\mathbbm{R}} \! \d v \, \e^{-v^2 / 2 C(0)} \left| v \right|^{p} 
      =  \Big(\frac{p-1}{2}\Big)! \, \frac{[2 C(0)]^{p/2}}{\pi^{1/2}},
    \end{equation}
    a Gaussian random potential is seen by Fubini's theorem to 
    satisfy (\ref{Eq:I+D}) and hence
    properties~\ass{S} and \ass{I}. 
    A simple sufficient criterion ensuring 
    $ \mathbbm{R}^d $-ergodicity, hence property \ass{E}, 
    is the mixing condition $\lim_{|x| \to \infty} C(x) = 0$.
    We note that the operator  $ H(A, V) $ is almost surely unbounded from below 
    for any Gaussian random potential $ V $.
  \end{nummer}
\end{remarks}
The next lemma has already been used to verify properties~\ass{S} and \ass{I} for the examples \ass{A} and \ass{P}.
It is patterned on \cite[Prop.~2]{KirMar83}, see also \cite[Cor.~V.3.4]{CaLa90}. 
\begin{lemma}\label{lem:mom}
  Let $( \Omega , \mathcal{A}, \mathbbm{P} ) $ be a probability space,
  $ \mu: \Omega \ni \omega \mapsto \mu^{(\omega)} $ be a random signed Borel measure on $ \mathbbm{R}^d $ and 
  $ u : \mathbbm{R}^d \to \mathbbm{R} $ be a Borel-measurable function. Let $ V $ be the random field given by
  the realizations 
  \begin{equation}\label{eq:Vallg}
    V^{(\omega)}(x) := \int_{\mathbbm{R}^d}  \mu^{(\omega)}( \d^d y) \,\, u(x-y), \qquad 
   x \in \mathbbm{R}^d , \quad\omega \in \Omega. 
   \end{equation}
   Then the estimate 
   \begin{align}\label{eq:Gewichse}
     & \Big\{ \mathbbm{E}\Big[\big( \int_{\Lambda(j)} \!\! \d^d x \, |V(x)|^{q} \big)^{r/q} \Big] \Big\}^{1/r} \notag \\
     & \quad \leq
     3^{d/q} \sup_{l \in \mathbbm{Z}^d} \Big\{ \mathbbm{E} \Big[ \big( | \mu | (\Lambda(l))\big)^r \Big] \Big\}^{1/r} \,
     \sum_{k \in \mathbbm{Z}^d} 
     \big(\int_{\Lambda(k)} \! \d^d x \, |u(x )|^{q} \big)^{1/q}
   \end{align}
   holds for all $ j \in \mathbbm{Z}^d $ and all those $ q, r\in [1,\infty[ $, for which the r.h.s.\ of (\ref{eq:Gewichse}) 
   is finite. 
   {\rm [}Here $ | \mu^{(\omega)} | $ denotes the total-variation measure of $ \mu^{(\omega)} $.{\rm ]}
\end{lemma}
\begin{proof}
  Minkowski's inequality \cite[Thm.~2.4]{LiLo97}, a subsequent shift in the $ \d^d x $-integration, and an enlargement of its domain show that
  \begin{align}
     \big( \int_{\Lambda(j)} \!\! \d^d x \, |V^{(\omega)}(x)|^{q} \big)^{1/q} 
     & \leq \int_{ \mathbbm{R}^d}  \big| \mu^{(\omega)} \big| ( \d^d y) \,\, 
                \big(\int_{\Lambda(j)} \! \d^d x \, |u(x-y )|^{q} \big)^{1/q} 
                \notag \\
     & \leq \sum_{k \in \mathbbm{Z}^d}  \big| \mu^{(\omega)} \big| (\Lambda(k)) \,\,
                 \big(\int_{\Lambda(j)-\Lambda(k)} \mkern-30mu  \d^d x \, |u(x )|^{q} \big)^{1/q},
   \end{align}
   where the cube $ \Lambda(j)-\Lambda(k) := \big\{ x - y \in \mathbbm{R}^d \, : 
   \, x \in \Lambda(j) \; \mbox{and} \;  y \in \Lambda(k) \big\} $ is the arithmetic 
   difference of the unit cubes $ \Lambda(j) $ and $ \Lambda(k) $.
   Using Minkowski's inequality again, we thus arrive at
   \begin{align}
      \Big\{ \mathbbm{E}\Big[\big( \int_{\Lambda(j)} \!\! \d^d x \, |V(x)|^{q} \big)^{r/q} \Big] \Big\}^{1/r}
        \leq \sum_{k \in \mathbbm{Z}^d} \Big\{ \mathbbm{E} \Big[ \big( | \mu | (\Lambda(k))\big)^r \Big] \Big\}^{1/r} \,
                \big(\int_{\Lambda(j)-\Lambda(k)} \mkern-30mu  \d^d x \, |u(x )|^{q} \big)^{1/q} & 
                \notag \\
        \leq  
              \sup_{l \in \mathbbm{Z}^d} \Big\{ \mathbbm{E} \Big[ \big( | \mu | (\Lambda(l))\big)^r \Big] \Big\}^{1/r} \,
              \sum_{k \in \mathbbm{Z}^d} \big(\int_{\Lambda(0)-\Lambda(0)} \mkern-30mu \d^d x \, |u(x-k)|^{q} \big)^{1/q}. &     
   \end{align}
   Since $ \Lambda(0)-\Lambda(0) $ is contained in the cube centered at the origin and consisting of $ 3^d $ unit cubes, 
   the proof is complete.
\end{proof}
                          %
%------------------------------------------------------------------------------
%------------------------------------------------------------------------------
\section{Proof of the Main Result}\label{AppIDOS}
The purpose of this section is to prove Lemma~\ref{Thm:DOSmass} and
Theorem~\ref{Thm:IDOS}, which is done in Subsection~\ref{Subsec:Proof}.
There we first 
show vague convergence of the density-of-states 
measures as claimed in Lemma~\ref{Thm:DOSmass}. 
Apart from minor modifications, we will thereto adapt the strategy of the 
proof of \cite[Thm.~5.20]{PaFi92} which presents an approximation argument
for the case $ A = 0 $. 
The latter permits us to take advantage of the independence of the infinite-volume 
limits of the boundary conditions
in case $V$ is bounded \cite{Nak00,DoIwMi01}. 
Moreover, the argument also allows us 
to use 
established results \cite{Uek94,BrHuLe93} 
in the case $ {\rm X} = {\rm D} $. 
This procedure requires auxiliary trace estimates, 
which are proven in Subsections~\ref{Subsec:Aux1} and \ref{Subsec:Aux2}.
In a second step, we use a criterion which provides conditions
under which vague convergence of measures implies
pointwise convergence of their
distribution functions.
This finally proves Theorem~\ref{Thm:IDOS}.
In Subsection~\ref{Subsec:VagueConv} we supply such a 
criterion and, to begin with, a criterion ensuring vague convergence. 

%%%%%%%%%%%%%%%%%%%%%%%%%%%%%%%%%%%%%%%%%%%%%%%%%%%%%%%%%%%%%%%%%%%%%%%%%%%%
%%%%%%%%%%%%%%%%%%%%%%%%%%%%%%%%%%%%%%%%%%%%%%%%%%%%%%%%%%%%%%%%%%%%%%%%%%%%

\subsection{On Vague Convergence of Positive Borel Measures on the Real Line}\label{Subsec:VagueConv}
We recall \cite[Def.~25.2(i)]{Bau92} that a positive measure on the real line $ \mathbbm{R} $
is a Borel measure if it assigns a finite length to each bounded 
Borel set in $ \mathbbm{R} $.
Note that every Borel measure on $ \mathbbm{R} $ is regular and hence
a Rad\'on measure \cite[Thm.~29.12]{Bau92}.
Moreover,
we recall from \cite[{${\S}$}~30, Exc.~3]{Bau92}
that vague convergence of a sequence 
of positive Borel measures $ (\mu_n)_{n \in \mathbbm{N}} $ on $ \mathbbm{R} $ to a measure $ \mu $
is equivalent to the convergence
$ \lim_{n \to \infty} \, \widehat\mu_n(f) =   \widehat\mu(f) $
for all $  f \in \mathcal{C}_0^1(\mathbbm{R}) $. Here and in the following, we occasionally use the
abbreviation
\begin{equation}
  \widehat\nu(f) := \int_{\mathbbm{R}} \nu(\d E)\, f(E)
\end{equation}
for the integral of a function $ f $ with respect to a measure $ \nu $.
 
Our proof of  Lemma~\ref{Thm:DOSmass} 
relies on the following generalization of \cite[Lemma~5.22]{PaFi92}
which provides a criterion for vague convergence. 
\begin{proposition}\label{Prop:Stieltjes}
  Let $ p \in ]1 , \infty [$ and let $\mu$ and $ \mu_n $, for each $ n \in \mathbbm{N} $, 
  be positive 
  (not necessarily finite) Borel measures on the real line
  $\mathbbm{R}$ such that the integrals
  \begin{equation}
    \widetilde \mu_n (z,p) := \int_{\mathbbm{R}} \frac{\mu_n(\d E)}{|E -z|^p} 
    \qquad \mbox{and}\qquad
    \widetilde \mu (z,p) := \int_{\mathbbm{R}} \frac{\mu(\d E)}{|E -z|^p}
  \end{equation}
  are finite for all $z \in \mathbbm{C}\backslash\mathbbm{R}$ and 
  all $n \in \mathbbm{N}$. 
  If ~$\lim_{n \to \infty} \widetilde \mu_n(z,p) =  \widetilde \mu(z,p) $~
  for all $z \in \mathbbm{C}\backslash\mathbbm{R}$,
  then $\mu_n$ converges vaguely to $\mu$ as $ n \to \infty$.
\end{proposition}
\begin{remark}
  The following implication is immediate.
  If $ \widetilde \mu (z,p) = \widetilde \nu (z,p) < \infty $ 
  for some $ p \in ]1 , \infty [ $ and all 
  $z \in \mathbbm{C}\backslash\mathbbm{R}$, then the underlying positive 
  Borel measures 
  $\mu $ and $\nu $ are equal.
\end{remark}
\begin{proof}[\bf Proof of Proposition~\ref{Prop:Stieltjes}]
  We first define the following one-parameter family 
  \begin{equation}\label{Eq:Approxdelta}
    E \mapsto \delta^{(\varepsilon)}_0(E) := \Upsilon_p \,\frac{\varepsilon^{p-1}}{\left|E -{\i}\varepsilon\right|^{p}},
    \qquad (\Upsilon_p)^{-1} := \int_{\mathbbm{R}} \! \frac{\!\d \xi}{|\xi - {\i} |^{p}}, \qquad \varepsilon > 0,
  \end{equation}
  of smooth Lebesgue probability densities on $ \mathbbm{R} $ which approximates 
  the Dirac measure $ \delta_0 $ on $ \mathbbm{R} $ supported
  at $ E = 0 $ as $ \varepsilon \downarrow 0 $. Moreover, let 
  $f_\varepsilon := \delta^{(\varepsilon)}_0 * f $ denote the convolution
  of $ \delta^{(\varepsilon)}_0 $ and $ f \in \mathcal{C}_0^1(\mathbbm{R}) $. 
  The fundamental theorem of calculus yields
  $ f(E - E') = f(E) - \int_0^{E'}\! \d\eta \, f'(E - \eta) $ and hence
  \begin{equation}
    \sup_{E \in \mathbbm{R}} \,  \left| f(E) - f_\varepsilon(E) \right| 
    \leq \int_{\mathbbm{R}} \! \d E' \,  \delta^{(1)}_0(E') \, \sup_{E \in \mathbbm{R}} \, 
     \biggl|\int_0^{\varepsilon E'} \! \d\eta \,  f'(E - \eta) \biggr|.
  \end{equation}
  The supremum on the r.h.s.\ does not exceed 
  $ \varepsilon \, \left| E' \right| \, \left| f' \right|_\infty $ and hence converges to zero
  as $  \varepsilon \downarrow 0 $ for all $ E' \in \mathbbm{R} $. 
  On the other hand, the supremum may 
  be estimated by $ \left| f' \right|_1 $ such that the dominated-convergence theorem is applicable and
  one has ~$ \lim_{ \varepsilon \downarrow 0 }  f_\varepsilon = f $ uniformly on $ \mathbbm{R} $.
  We now claim that there exists some 
  $C(\varepsilon) > 0$, depending on $ f $, with $ \lim_{\varepsilon \downarrow 0} C(\varepsilon) = 0 $ 
  such that
  \begin{equation}\label{Eq:unischranke}
    \left| f(E) - f_\varepsilon(E) \right| \leq C(\varepsilon)  \, \delta^{(1)}_0(E) 
  \end{equation}
  for all $E \in \mathbbm{R}$. To prove this, we pick a compact subset $ G $ of $ \mathbbm{R} $ such that 
  $\supp f \subset G $ and $ \dist (\mathbbm{R} \backslash G, \supp f) > 1 $. 
  Since $ f_\varepsilon $ converges uniformly to $f$ as $\varepsilon \downarrow 0 $, 
  the bound (\ref{Eq:unischranke}) is valid for all $ E \in G $. 
  For any other $ E \in \mathbbm{R} \backslash G $ the claim (\ref{Eq:unischranke}) follows from the estimate
  $ | f_\varepsilon(E) | \leq | f |_\infty \, \int_{\supp f}\d E' \, \delta^{(\varepsilon)}_0(E-E') $ and an explicit 
  computation.
  Inequality (\ref{Eq:unischranke}) may then be employed to show 
  \begin{equation}
    \int_{\mathbbm{R}}\!\mu_n(\d E)\, \left|  f(E) - f_\varepsilon(E) \right| \leq 
    C(\varepsilon) \,  \widetilde \mu_n(\,{\i}\,,p).
  \end{equation}
  The same holds true with $\mu$ and $\widetilde \mu$ taking the place of $\mu_n$ and $\widetilde \mu_n$, respectively. 
  We then estimate
  \begin{align}\label{Eq:e3}
    \big|  \widehat\mu_n(f) -  \widehat\mu(f) \big|
    & \leq \, 
     \big|   \widehat\mu_n(f) -  \widehat\mu_n(f_\varepsilon)  \big| 
    +   \big|   \widehat\mu(f) -  \widehat\mu(f_\varepsilon) \big| 
    +  \big|   \widehat\mu_n(f_\varepsilon) -  \widehat\mu(f_\varepsilon)  \big|  \notag  \\
    & \leq
    C(\varepsilon) \left( \widetilde \mu_n(\,{\i}\, , p) +  \widetilde \mu(\,{\i}\, , p) \right) 
    \notag \\
    & \quad 
    +   \Upsilon_p \varepsilon^{p-1} \!\! \int_{\mathbbm{R}}\! \d E \, f(E) \, 
    \left|  \widetilde \mu_n(E + {\i} \varepsilon , p) -  \widetilde \mu(E + {\i} \varepsilon , p) \right|.
    \quad
  \end{align}
  Here we have used the triangle inequality and
  Fubini's theorem in the integrals
  $ \widehat\mu_n(f_\varepsilon) = \int_\mathbbm{R}\mu_n(\d E)\, f_\varepsilon(E) $ and 
  $ \widehat\mu(f_\varepsilon) = \int_\mathbbm{R}\mu(\d E)\, f_\varepsilon(E) $. 
  The integral on the r.h.s.\ of (\ref{Eq:e3})
  tends to zero as $n \to \infty$
  by the dominated-convergence theorem. 
  It is applicable since the estimate  
  $ \widetilde\mu_n(E + {\i} \varepsilon, p ) \leq \left( 1 + \left| E \right|/\varepsilon \right)^p
  \widetilde\mu_n({\i} \varepsilon, p ) $
  shows that the integrand in (\ref{Eq:e3}) is bounded on $ \supp f $. Moreover, since 
  $  \widetilde\mu_n({\i} \varepsilon, p ) \to  \widetilde\mu({\i} \varepsilon, p ) $ as $ n \to \infty $,
  this bound may be chosen independent of $ n $.
  To complete the proof, we note that
  the other terms in (\ref{Eq:e3}) stay finite as $n \to \infty $ and
  can be made arbitrarily small as $\varepsilon \downarrow 0$.
\end{proof}
In case each term of a sequence $( \mu_n ) $ of measures possesses a 
finite (in general unbounded) distribution
function $E \mapsto \mu_n(]- \infty, E[) $, vague convergence of $  ( \mu_n ) $
does in general \emph{not} imply pointwise convergence of the sequence of distribution functions.
Even worse, if the latter convergence holds true, its limit is in general not equal to the distribution 
function of the limit of $  ( \mu_n ) $.
If one desires this equality,
one needs
a further criterion. This is provided by (\ref{Eq:tight}) in
\begin{proposition}\label{Prop:Kurt}
  Let ~$ \mu $ and $ \mu_n $, for each $n \in \mathbbm{N} $, be positive (not necessarily finite)
  Borel measures on the real line $ \mathbbm{R} $. Then the following two statements are equivalent:
  \begin{indentnummer*}
    \item
       The finiteness ~$ \mu\big( \left] - \infty , E \right[ \big) < \infty $~
       holds for all $  E \in \mathbbm{R} $ and the relation 
       \begin{equation}\label{eq:punktweise}
         \lim_{n \to \infty}  \, \mu_n\big( \left] - \infty , E \right[ \big) = \mu\big( \left] - \infty , E \right[ \big)
       \end{equation}
       holds for all $ E \in \mathbbm{R} $ except the at 
       most countably many with $ \mu(\{E\}) \neq 0$.
     \item The sequence $( \mu_n ) $ converges vaguely to $ \mu $ as $ n \to \infty $ and 
       the relation
       \begin{equation}\label{Eq:tight}
         \lim_{E \downarrow - \infty} \, \limsup_{n \to \infty} \,
         \mu_n\big( \left] - \infty , E \right[ \big) = 0  
       \end{equation}
       holds.
   \end{indentnummer*}
\end{proposition} 
\begin{remark}
      A sequence $( \mu_n ) $ obeying (\ref{Eq:tight}) might be called ``tight near minus infinity''.
      This naturally extends the usual notion of tightness \cite[{${\S}$}~30 Rem.~3]{Bau92}
      for finite measures to ones having only finite (in general unbounded) distribution functions and 
      ensures that no mass is lost at minus infinity as $ \mu_n $ tends to $ \mu $.
      More precisely, for each $ E \in \mathbbm{R} $ the sequence of truncated measures 
      $  ( \mu_n \,  I_E )_{n \in \mathbbm{N}} $, defined below (\ref{Def:Indfkt}), is tight in the usual sense.
      This follows either from the definition of the latter 
      or alternatively from the subsequent proof 
      and \cite[Thm.~30.8]{Bau92}.
\end{remark}
\begin{proof}[ \bf Proof of Proposition~\ref{Prop:Kurt}]
  \emph{(i) $ \Rightarrow $ (ii):} \hspace{0.1cm}
  Equation~(\ref{Eq:tight}) follows from (\ref{eq:punktweise}) and the finiteness 
  $ \mu\big( \left] - \infty , E \right[ \big) < \infty $.
  Moreover, for every $ f \in \mathcal{C}^1_0(\mathbbm{R}) $ one has  
  \begin{equation}
    \int_{\mathbbm{R}} \! \mu_n(\d E) \, f(E) 
    = -  \int_{\mathbbm{R}} \! \d E \, \mu_n(] - \infty , E [ ) \, f'(E)
  \end{equation}
  by partial integration. Vague convergence of $ (\mu_n ) $ to $ \mu $ is now a consequence of the 
  dominated-convergence theorem. It is applicable since (\ref{eq:punktweise}) implies the existence of
  a locally bounded function dominating all but finitely many of the non-decreasing functions 
  $ E \mapsto \mu_n\big( \left] - \infty , E \right[ \big) $.\\

  \noindent
  \emph{(ii) $ \Rightarrow $ (i):}  \hspace{0.1cm}
  For every $ E \in \mathbbm{R} $ we define
  the following continuous ``indicator function'' 
  \begin{equation}\label{Def:Indfkt}
    \mathbbm{R} \ni E' \mapsto I_E(E') 
    := \indfkt{] -\infty, E[}(E')  + \left(E + 1 - E' \right) \, 
    \indfkt{[E, E+1[}(E')
  \end{equation}
  of the half-line $ ] -\infty, E[ \, \subset \mathbbm{R} $. Moreover, we let $ \mu \,  I_E $
  denote the $ \mu $-continuous Borel measure with density $ I_E $, that is, 
  $ ( \mu \,  I_E ) (B) = \int_B \mu(\d E') \,  I_E(E') $ for all $ B \in \mathcal{B}(\mathbbm{R}) $,
  and the Borel measures $ \mu_n \,  I_E $ are defined accordingly.
  From~(\ref{Eq:tight}) it follows that  
  $  \limsup_{n \to \infty} \, \mu_n\big( \left] - \infty , E_0 \right[ \big) < \infty $
  for some $ E_0 \in \mathbbm{R} $.
  Hence the vague convergence $ \mu_n \,  I_E \to  \mu \,  I_E $ as $ n \to \infty $ and
  \cite[Lemma~30.3]{Bau92} imply that
  \begin{equation}
   \mu\big( \left] - \infty , E \right[ \big)   \leq   \widehat\mu( I_{E}) 
   \leq   \liminf_{n \to \infty} \,   \widehat\mu_n( I_{E}) 
   \leq  \liminf_{n \to \infty} \,   \widehat\mu_n\big( ] - \infty , E+1 [ \big) < \infty \qquad
   \end{equation}
   for all $ E \in ] - \infty, E_0 -1 ] $.
   Since $ \mu $ is a Borel measure, this implies 
   the finiteness of ~$  \mu\big( \left] - \infty , E \right[ \big) = 
   \mu\big( \left] - \infty , E_0 - 1 \right[ \big) +  \mu\big( \left[  E_0 - 1, E \right[ \big)$ 
   for all $ E \in \mathbbm{R} $.
   
   The sequence of
   total masses of ~$ \mu_n \, I_E  $
   converges to the total mass of the limiting measure ~$ \mu \,  I_E  $. 
   More precisely, defining the function ~$ J_{E_1,E} := I_E - I_{E_1}  \in \mathcal{C}_0(\mathbbm{R}) $~
   for each $ E_1 < E $, it follows that
   \begin{align}
     \lim_{n \to \infty} \,   \widehat\mu_n(I_E) 
      & = \lim_{E_1 \downarrow - \infty} \,  \lim_{n \to \infty} \,
       \widehat\mu_n( I_{E_1}) +  
      \lim_{E_1 \downarrow - \infty} \, \lim_{n \to \infty} \,  \widehat\mu_n(J_{E_1,E}) \notag \\    
     & =  \lim_{E_1 \downarrow - \infty}  \widehat\mu(J_{E_1,E}) 
     =  \widehat\mu(I_E).
   \end{align}
   Here the first term on the r.h.s.\ of the first equality tends to zero using (\ref{Eq:tight})
   and $ 0 \leq \widehat\mu_n(I_{E_1}) \leq \mu_n(] - \infty , E_1 +1 [) $. 
   The second equality is
   a consequence of the vague convergence of $ \mu_n $ to $ \mu $ as $ n \to \infty $.
   The third equality follows from the monotone-convergence theorem. 
   Hence \cite[Thm.~30.8]{Bau92}
   implies that ~$ \mu_n \,  I_E $ 
   converges weakly to ~$ \mu \, I_E $ as 
   $ n \to \infty $, not only vaguely.
   We recall from \cite[Def.~30.7]{Bau92} that weak convergence of the latter sequence requires
   that $ \widehat\mu_n(I_E f) $ tends to $ \widehat\mu(I_E f) $ as $ n \to \infty $ for every
   bounded continuous function $ f $.
   The claimed convergence (\ref{eq:punktweise}) of the corresponding distribution functions 
   is therefore reduced to the content of \cite[Thm.~30.12]{Bau92}.
\end{proof}

%%%%%%%%%%%%%%%%%%%%%%%%%%%%%%%%%%%%%%%%%%%%%%%%%%%%%%%%%%%%%%%%%%%%%%%%%%%

\subsection{Proofs of Lemma~\ref{Thm:DOSmass} and Theorem~\ref{Thm:IDOS}}\label{Subsec:Proof}
%\subsection{Proofs of Lemma~3.1 and Theorem~3.1}\label{Subsec:Proof}
%
We first give a
\begin{proof}[ \bf Proof of  Lemma~\ref{Thm:DOSmass}]
To show that $ \nu $ is a
positive Borel measure on $\mathbbm{R}$, it suffices 
that
\begin{align}\label{Eq:DOSmassgibts}
  \nu(I) \,  \left| \Gamma \right| & = \mathbbm{E}\Big\{
    \Tr\Big[ \indfkt{\Gamma} \, \indfkt{I}\big(H(A,V)\big)  \, \indfkt{\Gamma} \big]\Big\}
  \notag \\
  & \leq \left(\sqrt{2} \varepsilon
  \right)^{2\vartheta}  \mathbbm{E}\Big\{
    \Tr\big[ \indfkt{\Gamma} \, | H(A,V) - E_0 - {\i} \varepsilon |^{-2\vartheta} \, 
      \indfkt{\Gamma} \big]\Big\} < \infty \quad
\end{align}
for any compact energy interval $I=\left[E_0-\varepsilon , E_0+\varepsilon \right]$, 
$ E_0 \in \mathbbm{R} $, $\varepsilon > 0$. 
This follows from the elementary inequality 
$ \indfkt{I}(E) \leq (\sqrt{2} \varepsilon )^{2\vartheta} \left| E - E_0 - {\i} \varepsilon \right|^{-2\vartheta} $,
the spectral theorem applied to $ H(A,V^{(\omega)}) $ and the functional calculus.
Proposition~\ref{Eq:pore21} below and property~\ass{I}
ensure that the r.h.s.\ of (\ref{Eq:DOSmassgibts}) is indeed finite.

To prove (\ref{Eq:DOSvag}) 
we employ an approximation argument with bounded truncated random potentials given by
\begin{equation}\label{Def:Vn}
    V_n^{(\omega)}(x):= V^{(\omega)}(x) \; \, 
    \Theta\left(n - |V^{(\omega)}(x)|\right), \quad n \in \mathbbm{N}.
\end{equation}
We denote by $\nu^{(\omega)}_{\Lambda,{\rm X},n}$, 
with $ \rm X = \rm D $ or $ \rm X = \rm N $, the approximate finite-volume 
density-of-states measure associated with $ V_n $, see (\ref{Def:nu}).
Moreover,
\begin{equation}
    \nu_n(I):= \frac{1}{|\Gamma|} \, 
    \mathbbm{E}\Big\{ \Tr\big[\indfkt{\Gamma}\indfkt{I}\big(H(A,V_n)\big) 
        \indfkt{\Gamma}\big]\Big\}, \quad  I \in \mathcal{B}(\mathbbm{R}),
\end{equation}
defines the approximate (infinite-volume) density-of-states measure. It is a
positive Borel measure on $ \mathbbm{R} $, see (\ref{Eq:DOSmassgibts}), 
and independent of the bounded open
cube $\Gamma \subset \mathbbm{R}^d $ due to $\mathbbm{Z}^d$-homogeneity.
In case $ {\rm X } = {\rm D} $ we let $ f \in \mathcal{C}_0^1(\mathbbm{R}) $ and estimate 
as follows 
\begin{multline}\label{eq:epsdrittel}
  \big| \, \left| \Lambda \right|^{-1}\, \widehat\nu^{(\omega)}_{\Lambda, {\rm D}}(f) -
  \widehat\nu(f)  \, \big| 
  \leq  \big| \,   \widehat\nu_n(f) -  \widehat\nu(f)  \, \big| \\
  +
  \left| \Lambda \right|^{-1} \, \big| \,  
  \widehat\nu^{(\omega)}_{\Lambda,{\rm D},n}(f) -  \widehat\nu^{(\omega)}_{\Lambda,{\rm D}}(f)
  \, \big|   + 
  \big| \, \left| \Lambda \right|^{-1}  \widehat\nu^{(\omega)}_{\Lambda,{\rm D},n}(f) -  \widehat\nu_n(f)  \, \big|.
  \quad
\end{multline}
We first consider the limit $ \Lambda \uparrow \mathbbm{R}^d $. In this limit, the third difference on
the r.h.s.\ of (\ref{eq:epsdrittel}) vanishes for all 
$ \omega \in \widehat \Omega := \bigcap_{n \in \mathbbm{N}} \Omega_{n} $
and all $ n \in \mathbbm{N} $ by Lemma~\ref{LemmaB1} below.
Next we consider the limit $ n \to \infty $, in which
the second difference vanishes for all $ \omega \in \widetilde \Omega $
by Lemma~\ref{LemmaB2}. In the latter limit, the first difference vanishes by Lemma~\ref{LemmaB3}.
This proves the claimed vague convergence of $ | \Lambda |^{- 1} \nu_{\Lambda, {\rm D}}^{(\omega)} $ to
$ \nu $ as $ \Lambda \uparrow \mathbbm{R}^d $ for all $ \omega \in \widehat \Omega \cap \widetilde \Omega $, 
hence for $ \mathbbm{P} $-almost all $ \omega \in \Omega $. 
In case $ {\rm X } = {\rm N} $ we estimate 
\begin{equation} 
   \big| \, \left| \Lambda \right|^{-1}\, \widehat\nu^{(\omega)}_{\Lambda, {\rm N}}(f) -
  \widehat\nu(f)  \, \big| 
  \leq  \big| \, \left| \Lambda \right|^{-1}\, \widehat\nu^{(\omega)}_{\Lambda, {\rm D}}(f) -
  \widehat\nu(f)  \, \big| 
  + \left| \Lambda \right|^{-1} \,
  \big| \, \widehat\nu^{(\omega)}_{\Lambda, {\rm N}}(f) - \widehat\nu^{(\omega)}_{\Lambda, {\rm D}}(f) \, \big|. 
\end{equation}
As $ \Lambda \uparrow \mathbbm{R}^d $ the first term on the r.h.s.\ converges to zero 
for $\mathbbm{P}$-almost all $ \omega \in \Omega $ and the same is true for the second term 
thanks to Proposition~\ref{Prop:B1} below. 
\end{proof}
We now prove our main result. 
\begin{proof}[ \bf Proof of Theorem~\ref{Thm:IDOS}]
  Since we have already established 
  the vague convergence of the density-of-states measures in Lemma~\ref{Thm:DOSmass},
  it remains to verify relation~(\ref{Eq:tight}) of Proposition~\ref{Prop:Kurt}
  for the corresponding random distribution functions 
  $  \left| \Lambda \right|^{-1} \,  N_{\Lambda, \rm X}^{(\omega)} $~
  for $ \mathbbm{P} $-almost all $ \omega \in \Omega $.
  
  To this end, we employ the elementary inequality 
  $ \Theta(E) \leq 2 \, \int_{- \infty}^E \! \d E' \, \delta^{(\varepsilon)}_0(E') $
  valid for all $ E \in \mathbbm{R} $, $ \varepsilon > 0 $ with $\delta^{(\varepsilon)}_0 $  
  defined in (\ref{Eq:Approxdelta}). 
  Choosing $ \varepsilon = 1 $ and $ p = 2 \vartheta + 1 $ there, 
  we get
  \begin{align}
    N_{\Lambda, \rm X}^{(\omega)}(E)  & = \Tr \Big[\Theta\big( E - H_{\Lambda, \rm X}(A,V^{(\omega)})\big)\Big] 
    \notag \\
    & \leq
    2 \, \Upsilon_{2\vartheta+1} \, \int_{- \infty}^E \! \d E' \,\, \,  
    \widetilde \nu_{\Lambda, \rm X}^{(\omega)}(E' - {\i} ,2\vartheta+1)
    \notag \\
    &  \leq 4 \, \left| E \right|^{\frac{d}{2} - 2\vartheta} \,
     \frac{\Upsilon_{2\vartheta+1}}{4\vartheta -d} \,\, C_1(1) \,
   \int_{\Lambda} \! \d^d x \,\, \big( 2  + |V^{(\omega)}(x)| \big)^{2\vartheta+1}
   \label{eq:IDOSEndl}
  \end{align}
  for all $ E \in ] - \infty , -1 ] $. Here, the second inequality results from (\ref{eq:endlvol1}) and
  Proposition~\ref{Eq1:Prop1} choosing $ E_1 = E' $ there.
  Dividing (\ref{eq:IDOSEndl}) by the volume $ \left| \Lambda \right| $ and 
  using the Birkhoff-Khintchine ergodic theorem in the formulation \cite[Prop.~1.13]{PaFi92} we get
  \begin{equation}\label{eq:IDOSEndl2}
    \limsup_{ \Lambda \uparrow \mathbbm{R}^d } \, 
    \frac{ N_{\Lambda, \rm X}^{(\omega)}(E) }{ \left| \Lambda \right|} \leq 
      4 \, \left| E \right|^{\frac{d}{2} - 2\vartheta} \,
     \frac{\Upsilon_{2\vartheta+1}}{4\vartheta -d} \,\, C_1(1) \,\,
     \mathbbm{E}\Big[  \int_{\Lambda(0)} \! \d^d x \,\, \big( 2  + |V(x)| \big)^{2\vartheta+1} \Big]
  \end{equation}
  for $ \mathbbm{P} $-almost all $ \omega \in \Omega $ and all $ E \in ] - \infty , - 1 ] $. 
  Since the r.h.s.\ of (\ref{eq:IDOSEndl2}) converges to zero 
  as $ E \downarrow - \infty $, relation~(\ref{Eq:tight})
  of Proposition~\ref{Prop:Kurt} is fulfilled.
  The existence of the distribution function $ N(E) = \nu(] - \infty , E [) $ of the limiting measure 
  $ \nu $ for all $ E \in \mathbbm{R} $
  as well as the claimed convergence are thus warranted by Proposition~\ref{Prop:Kurt}. 
\end{proof}
%%%%%%%%%%%%%%%%%%%%%%%%%%%%%%%%%%%%%%%%%%%%%%%%%%%%%%%%%%%%%%%%%%%%%%%%%%%
%%%%%%%%%%%%%%%%%%%%%%%%%%%%%%%%%%%%%%%%%%%%%%%%%%%%%%%%%%%%%%%%%%%%%%%%%%%%%%%%%%%%%%%%%%%%%%%%%%%%%%%%%%%
%
The proof of Lemma~\ref{Thm:DOSmass} was based on three lemmas and two propositions.
The first lemma basically recalls known facts \cite{Uek94,BrHuLe93} for $ {\rm X} = {\rm D} $ and
$V$ bounded.
\begin{lemma}\label{LemmaB1}
  Let $ \Lambda \subset \mathbbm{R}^d $ stand for bounded open cubes.
  Suppose $ A $ and $ V $ have the properties \ass{C} and \ass{E}. Then
  for every $n \in \mathbbm{N}$ there exists $ \Omega_n \in \mathcal{A} $ with 
  $ \mathbbm{P}(\Omega_n) = 1 $
  such that 
  \begin{equation}
    \lim_{ \Lambda \uparrow \mathbbm{R}^d} \, 
    \frac{\nu^{(\omega)}_{\Lambda,{\rm D},n}}{\left| \Lambda \right|} = \nu_n
  \end{equation}
  vaguely for all $ \omega \in  \Omega_n $.
\end{lemma}
\begin{proof} 
  See \cite[Thm.~3.1 and Prop.~3.1(ii)]{Uek94} where 
  the appropriate Feynman-Kac-It{\^o} formula for the infinite-volume and 
  the Dirichlet-finite-volume Schr{\"o}dinger semigroup is employed; 
  see also \cite{BrHuLe93}.
\end{proof}
%%%%%%%%%%%%%%%%%%%%%%%%%%%%%%%%%%%%%%%%%%%%%%%%%%%%%%%%%%%%%%%%%%%%%%%%%%%%%%%%%%%%
%
\begin{lemma} \label{LemmaB2}
  Let $ \Lambda \subset \mathbbm{R}^d $ stand for bounded open cubes. Let
  $ A $ and $ V $ be supplied with the properties~\ass{C}, \ass{I}, and \ass{E}.
  Then there exists  $ \widetilde \Omega\in \mathcal{A} $ with $ \mathbbm{P}( \widetilde \Omega) = 1 $
  such that
  \begin{equation}\label{Eq:LemmaB2}
    \lim_{n \to \infty} \limsup_{\Lambda \uparrow \mathbbm{R}^d} \, 
    \frac{1}{\left| \Lambda \right|} \,
    \big| \,  
    \widehat\nu^{(\omega)}_{\Lambda,{\rm X},n}(f) -  \widehat\nu^{(\omega)}_{\Lambda,{\rm X}}(f)
    \, \big|
    = 0 
  \end{equation}
  for all $f \in \mathcal{C}_0^1(\mathbbm{R}) $, all $ \omega \in  \widetilde \Omega $
  and both boundary conditions ${\rm X}= {\rm D}$ and ${\rm X}={\rm N}$.
\end{lemma}
\begin{proof}
  Thanks to (\ref{eq:endlvol1}), Proposition~\ref{Eq1:Prop1} below and property~\ass{I}, the integrals
  \begin{equation}\label{eq:nutildeendlvol}
     \widetilde \nu^{(\omega)}_{\Lambda,{\rm X},n}(z,2\vartheta) = 
     \int_{\mathbbm{R}} \frac{\nu^{(\omega)}_{\Lambda,{\rm X},n}(\d E)}{ |E -z|^{2\vartheta}} 
     =  
     \Tr\Big[ \big| H_{\Lambda, {\rm X}}(A, V^{(\omega)}_{n}) - z \big|^{-2\vartheta} \Big]
  \end{equation}
  and (analogously) $ \widetilde \nu^{(\omega)}_{\Lambda,{\rm X}}(z,2\vartheta) $
  are finite for all $z \in \mathbbm{C}\backslash\mathbbm{R}$ and 
  $\mathbbm{P}$-almost all $\omega \in \Omega$ such that (\ref{Eq:e3}) yields
  \begin{multline}\label{Eq:Folgerunge3} 
    \big| \, \widehat\nu^{(\omega)}_{\Lambda,{\rm X},n}(f) - 
    \widehat\nu^{(\omega)}_{\Lambda,{\rm X}} (f) \, \big|
    \leq C(\varepsilon) \left[ \, \widetilde \nu^{(\omega)}_{\Lambda,{\rm X},n}(\,{\i} ,2\vartheta)
      + \widetilde \nu^{(\omega)}_{\Lambda,{\rm X}}(\,{\i} ,2\vartheta) \right] \\ 
    + \Upsilon_{2\vartheta}  \, \varepsilon^{2\vartheta-1} \, | f |_1 \, 
    \sup_{E \in \supp f} \big|   \widetilde \nu^{(\omega)}_{\Lambda,{\rm X},n}(E + {\i} \varepsilon ,2\vartheta)
      - \widetilde \nu^{(\omega)}_{\Lambda,{\rm X}}(E + {\i} \varepsilon ,2\vartheta) \big|.
  \end{multline}
  Here the quantity $ C(\varepsilon) $, which depends on $ \varepsilon $ and $ f $, 
  was introduced in (\ref{Eq:unischranke}) and vanishes for $ \varepsilon \downarrow 0 $.
  We further estimate the first term with the help of (\ref{eq:endlvol1}) and 
  Proposition~\ref{Eq1:Prop1} choosing $ E_1 = -1 $
  there. The upper limit $ \Lambda \uparrow \mathbbm{R}^d $ of the first term after dividing
  by the volume $ \left| \Lambda \right| $ is then
  seen to be finite by the Birkhoff-Khintchine ergodic theorem \cite[Prop.~1.13]{PaFi92},
  \begin{equation}
    \limsup_{\Lambda \uparrow \mathbbm{R}^d} \,  \frac{1}{\left| \Lambda \right|} \, \left[ 
      \widetilde \nu^{(\omega)}_{\Lambda,{\rm X},n}(\,{\i} ,2\vartheta)
      + \widetilde \nu^{(\omega)}_{\Lambda,{\rm X}}(\,{\i} ,2\vartheta) \right] 
    \leq
    2 \, C_1(1) \,  \mathbbm{E}\Big[ \int_{\Lambda(0)} \!\! \d^d x \,
      \big(3 + |V(x)| \big)^{2\vartheta} \Big]
    \quad
  \end{equation}
  for $\mathbbm{P}$-almost all $\omega \in \Omega $. The second term in (\ref{Eq:Folgerunge3}) 
  is bounded with the help of (\ref{eq:endlvol2})
  and Proposition~\ref{Eq2:Prop1} 
  where we again choose $ E_1 = -1 $.
  This bound together with the same ergodic theorem yields 
  \begin{multline}\label{Eq:FolgErgB}
     \limsup_{\Lambda \uparrow \mathbbm{R}^d} \,  
     \sup_{E \in \supp f}  \,   \frac{1}{\left| \Lambda \right|} \,  \big| \,
     \widetilde \nu^{(\omega)}_{\Lambda,{\rm X},n}(E + {\i} \varepsilon ,2\vartheta)
      - \widetilde \nu^{(\omega)}_{\Lambda,{\rm X}}(E + {\i} \varepsilon ,2\vartheta) \big| \\
     \leq
     C_2(\varepsilon)  \, 
     \left\{ \mathbbm{E}\Big[ \int_{\Lambda(0)} \!\! \d^d x \, 
       \big(2 + \sup_{E \in \supp f} |E + {\i} \varepsilon |  
       + |V(x)| \big)^{2\vartheta+1} \Big] \right\}^{\frac{2\vartheta}{2\vartheta+1}} \\
     \times
     \left\{ \mathbbm{E}\Big[ \int_{\Lambda(0)} \!\! \d^d x \, \big| V(x) - V_n(x) \big|^{2\vartheta+1}
       \Big] \right\}^{\frac{1}{2\vartheta+1}}
  \end{multline}
  for $\mathbbm{P}$-almost all $\omega \in \Omega$. In the limit $n \to \infty$, 
  the r.h.s.\ and hence the l.h.s.\ of (\ref{Eq:FolgErgB}) 
  vanishes for $\mathbbm{P}$-almost all $\omega\in \Omega$ thanks to property~\ass{I}. 
  This completes the proof since the
  first term on the l.h.s.\ of (\ref{Eq:Folgerunge3})
  may be made arbitrarily small as $\varepsilon \downarrow 0$.
\end{proof}
%%%%%%%%%%%%%%%%%%%%%%%%%%%%%%%%%%%%%%%%%%%%%%%%%%%%%%%%%%%%%%%%%%%%%%%%%%%%%%
%
The last lemma shows in which sense the approximate (infinite-volume)
density-of-states measures approach the exact one.
\begin{lemma}\label{LemmaB3}
  Suppose $A $ and $ V $ have the properties \ass{C}, \ass{S}, \ass{I}, and \ass{E}. 
  Then $ \nu_n $ converges vaguely to $\nu$ as
  $n \to \infty$.
\end{lemma}
\begin{proof}
  Thanks to Proposition~\ref{Eq:pore21} and property~\ass{I}, the
  integrals
  \begin{equation}\label{Def:nutilde}
    \widetilde \nu_n(z,2\vartheta) = \int_{\mathbbm{R}} \frac{\nu_{n}(\d E)}{ |E -z|^{2\vartheta}} 
    = \frac{1}{|\Gamma|} \, \mathbbm{E}\left\{ \Tr \left[
        \indfkt{\Gamma} 
        \left| H(A,V_n) -z \right|^{-2\vartheta}
        \indfkt{\Gamma} \right] \right\}
  \end{equation}
  and (analogously) $ \widetilde \nu (z,2\vartheta) $ are finite for all $z \in
  \mathbbm{C}\backslash\mathbbm{R}$.  Moreover, $\lim_{n \to \infty}
  \widetilde \nu_n(z,2\vartheta) = \widetilde \nu(z,2\vartheta)$ for all $z \in
  \mathbbm{C}\backslash\mathbbm{R}$ by (\ref{eq:infinitevol}),
  Proposition~\ref{Eq:pore22} and property~\ass{I} again. This implies vague convergence by
  Proposition~\ref{Prop:Stieltjes}.  
\end{proof}
%%%%%%%%%%%%%%%%%%%%%%%%%%%%%%%%%%%%%%%%%%%%%%%%%%%%%%%%%%%%%%%%%%%%%%%%%%%
In the following proposition we exploit recent results of Nakamura \cite{Nak00} or 
Doi, Iwatsuka and Mine \cite{DoIwMi01}
on the independence of the density-of-states measure of the chosen boundary
condition for the present setting, thereby heavily relying on either of these results.
\begin{proposition}\label{Prop:B1}
  Let $ \Lambda \subset \mathbbm{R}^d $ stand for bounded open cubes. Assume
  $ A $ is a vector potential with property~\ass{C} and $ V $ is a
  random potential with properties~\ass{I} and \ass{E}. 
  Then
  \begin{equation}
    \lim_{\Lambda \uparrow \mathbbm{R}^d} \, 
    \frac{1}{\left| \Lambda \right| } \, 
     \big| \int_{\mathbbm{R}}\!\! \nu^{(\omega)}_{\Lambda,{\rm N}}(\d E) \, f(E) - 
    \int_{\mathbbm{R}}\!\! \nu^{(\omega)}_{\Lambda,{\rm D}} (\d E) \, f(E) \, \big|
    = 0
  \end{equation}
  for all $ f
  \in \mathcal{C}^1_0(\mathbbm{R}) $ and all $ \omega\in  \widetilde \Omega$. 
  {\rm [}The set $ \widetilde \Omega$ is defined
  in Lemma~\ref{LemmaB2}.{\rm ]}
\end{proposition}
\begin{remark}
  An extension of Nakamura's result \cite[Thm.~1]{Nak00} (without his error estimate) 
  to
  unbounded \emph{non-random} potentials $ v $ may be achieved with the subsequent 
  techniques under the uniform local integrability condition 
  $ \sup_{y \in \mathbbm{Z}^d} \int_{\Lambda(y)} \d^d x \, \left| v (x) \right|^{2 \vartheta +1} < \infty $,
  where $ \vartheta $ is the smallest integer with $ \vartheta > d/4 $.
  Properties~\ass{I} and \ass{E} of a random potential in general 
  do not imply this condition $ \mathbbm{P} $-almost surely.
\end{remark}
%%%%%%%%%%%%%%%%%%%%%%%%%%%%%%%%%%%%%%%%%%%%%%%%%%%%%%% 
\begin{proof}[ \bf Proof of Proposition~\ref{Prop:B1}]
  The proof consists of an approximation argument.  To this end, we
  recall the definition~(\ref{Def:Vn}) of the truncated random potential $ V_n $.
  Since $ V_n $ is
  bounded and $ A $ enjoys property~\ass{C}, 
  we may apply \cite[Thm.~1]{Nak00} (or \cite[Thm.~1.2]{DoIwMi01} together with Lemma~\ref{LemmaB1}) which gives 
   \begin{equation}\label{eq:Naka}
    \lim_{\Lambda \uparrow \mathbbm{R}^d} \, 
    \frac{1}{\left| \Lambda \right| } \, \left| 
    \Tr \left[f(H_{\Lambda, {\rm N}}(A,
    V_n^{(\omega)})) - f(H_{\Lambda, {\rm D}}(A,
    V_n^{(\omega)})) \right]
     \, \right| 
    = 0
  \end{equation}
  for all  $ n \in \mathbbm{N} $, all $ f
  \in \mathcal{C}^1_0(\mathbbm{R}) $ and all $ \omega \in \Omega $.
  Using the triangle inequality we estimate
  \begin{equation}
    \big| \widehat\nu_{\Lambda, \rm N}^{(\omega)}(f) - \widehat\nu_{\Lambda, \rm D}^{(\omega)}(f) \big| 
    \leq \big| \widehat\nu_{\Lambda, {\rm N}, n}^{(\omega)}(f) - 
     \widehat\nu_{\Lambda, {\rm D}, n}^{(\omega)}(f) \big| 
     +
    \sum_{\rm X = \rm D, \rm N} \, \big| 
    \widehat\nu_{\Lambda, {\rm X}, n}^{(\omega)}(f) - \widehat\nu_{\Lambda, \rm X}^{(\omega)}(f)
     \big|.
     \quad
  \end{equation}
  The proof is then completed
  with the help of (\ref{eq:Naka}) and Lemma~\ref{LemmaB2}. 
  %which
%  controls the difference of the approximate
%  and the exact finite-volume
%  density-of-states measures 
%  $\nu^{(\omega)}_{\Lambda,{\rm X},n}$ and $
%  \nu^{(\omega)}_{\Lambda,{\rm X}}$, respectively.
%
\end{proof}
%
%%%%%%%%%%%%%%%%%%%%%%%%%%%%%%%%%%%%%%%%%%%%%%%%%%%%%%%%%%%%%%%%%%%%%%%%%%%%

Various proofs in the present subsection 
rely on estimates stated in Proposition~\ref{Prop:pore1} and Proposition~\ref{Prop:pore2}.
These propositions will be proven in the remaining two subsections. 
In fact, they extend parts of Lemma~5.4 (resp. 5.7) and
Lemma~5.12 (resp.  5.14) in \cite{PaFi92} to the case of non-zero vector potentials.
Basically, the extensions follow from the so-called diamagnetic inequality. For this inequality
the reader may find useful the compilation \cite[App.~A.2]{HuLeMuWa01} which covers the 
Neumann-boundary-condition case $ \rm X = \rm N $.
\subsection{Finite-volume trace-ideal estimates}\label{Subsec:Aux1} 
%
%Throughout this subsection we will use the notations 
%\begin{equation}\label{eq:notation}
%  \left| f \right|_{p} := 
%  \Big(  \int_{\Lambda} \! \d^d x \, \left| f(x) \right|^p \Big)^{\frac{1}{p}}, \quad 
%   \left\| F \right\|_p := \Big( \Tr \left| F \right|^{p} \Big)^{\frac{1}{p}}, \quad
%   \left| F \right| := \left( F^\dagger F \right)^{\frac{1}{2}},
%\end{equation}
%with $  p \in [1 , \infty[ $ and open $ \Lambda \subseteq \mathbbm{R}^d $, 
%for the norm of a function $f \in {\rm L}^p(\Lambda)$, the 
%(von Neumann-) Schatten norm of an operator $ F $ on $ {\rm L}^2(\Lambda) $ in the Banach space 
%$ \mathcal{J}_p\left({\rm L}^2(\Lambda)\right) $ and the modulus of a closed operator $F$, respectively.
%For the $  \mathcal{J}_p $-spaces of compact operators, see \cite{Sim79TR,BirSol87}.\\

Our first aim is to estimate the trace norm  $ \| \cdot \|_1 $ 
(recall the notation~(\ref{eq:notation})) of a power of the resolvent of the finite-volume 
magnetic Schr{\"o}dinger operator $ H_{\Lambda, \rm X}(a,v) $ and of the difference of
two such powers.
%
%%%%%%%%%%%%%%%%%%%%%%%%%%%%%%%%%%%%%%%%%%%%%%%%%%%%%%%%%%%%%%%%%%%%%%%%%%%%%%%%%%%%%%%%%%%%%%%%
%%%%%%%%%%%%%%%%%%%%%%% PROPOSITION FINITE VOLUME %%%%%%%%%%%%%%%%%%%%%%%%%%%%%%%%%%%%%%%%%%%%%%
%%%%%%%%%%%%%%%%%%%%%%%%%%%%%%%%%%%%%%%%%%%%%%%%%%%%%%%%%%%%%%%%%%%%%%%%%%%%%%%%%%%%%%%%%%%%%%%%
%
\begin{proposition}\label{Prop:pore1}
  Let $\Lambda \subset \mathbbm{R}^d $ be a bounded open cube with $
  \left| \Lambda \right| \geq 1 $ and  ~$ \rm X = \rm D $ or
  $ \rm X =\rm N $. Let $k \in \mathbbm{N} $ with $ k > d/4$
  and $ E_1 \in ] -\infty , -1 ]$.  
  Let ~$ a $~ be a vector potential
  with $ \left| a \right|^2 \in {\rm L}^1_{\rm loc}(\mathbbm{R}^d) $
  and $ v, v' \in {\rm L}^{2k + 1}_{\rm loc}(\mathbbm{R}^d)$ be two
  scalar potentials with $ |v'| \leq |v| $.  
  Then for every $ z \in \mathbbm{C}\backslash\mathbbm{R} $ there exist 
  two constants $ C_1(\Im z) $,  $C_2(\Im z) > 0 $, which depend on $ d $ and $ p $, but
  are independent of
  $\Lambda$, $ \rm X $, $a$, $v$, $v'$ and $ E_1 $, such that
  \begin{nummer}
    \item\label{Eq1:Prop1}
    ~~$ \displaystyle
    \big\| \left( H_{\Lambda, {\rm X}}(a, v) - z \right)^{-p} \big\|_{1}
    \leq C_1(\Im z) \, \left| E_1 \right|^{\frac{d}{2}-p}
    \,  \big| 1 + |z-E_1| + |v| \big|_{p}^p $\\[2ex] 
    \hspace*{1cm} for all $ p \in [2,  2k + 1] \, \cap \, ] \, d/2 , 2k+1 ]$,
    \item\label{Eq2:Prop1}
    ~~$ \displaystyle  
    \big\| \left( H_{\Lambda, {\rm X}}(a, v) - z \right)^{-2k} - 
    \left( H_{\Lambda, {\rm X}}(a, v') - z \right)^{-2k} \big\|_1 $\\[2ex] 
     \hspace*{2cm} $ \displaystyle 
    \leq C_2(\Im z) \, \left| E_1 \right|^{\frac{d}{2}-2k} \, \big| 1 + |z-E_1| + |v| \big|_{2k+1}^{2k} \, 
    \big| v- v' \big|_{2k+1}$.
  \end{nummer}
\end{proposition}
\begin{remarks}
  \begin{nummer}
    \item
      We recall from \cite[App.]{HuLeMuWa01} that the assumptions of Proposition~\ref{Prop:pore1} guarantee
      that the operators 
      $ H_{\Lambda, {\rm X}}(a,v)$ for $ \rm X = \rm D $
      and $ \rm X =  \rm N $ are well defined as self-adjoint operators via forms.
    \item
      From (\ref{eq:nutildeendlvol}) we conclude that
      \begin{equation}\label{eq:endlvol1}
      \widetilde \nu_{\Lambda, {\rm X}}^{(\omega)}(z, p) = 
      \big\| ( H_{\Lambda, {\rm X}}(A, V^{(\omega)}) - z )^{-p}  \big\|_1,
      \end{equation}
      because the resolvent of $ H_{\Lambda, {\rm X}}(A, V^{(\omega)}) $ commutes with its adjoint.
      Therefore, Proposition~\ref{Prop:pore1} provides upper bounds 
      on $\widetilde \nu_{\Lambda, {\rm X}}^{(\omega)}(z, 2\vartheta) $,
      $\widetilde \nu_{\Lambda, {\rm X},n}^{(\omega)}(z, 2\vartheta)$,  
      and $ \widetilde \nu_{\Lambda, {\rm X}}^{(\omega)}(z, 2\vartheta+1) $ 
      as well as on the r.h.s.\ of the estimate
      \begin{align}\label{eq:endlvol2}
         & \left| \, \widetilde \nu_{\Lambda, {\rm X}}^{(\omega)}(z, 2\vartheta) - 
          \widetilde\nu_{\Lambda, {\rm X},n}^{(\omega)}(z, 2\vartheta) \right| \notag \\  
         & \quad \leq \Big\| \big( H_{\Lambda, {\rm X}}(A, V^{(\omega)}) - z \big)^{-2\vartheta} -
          \big( H_{\Lambda, {\rm X}}(A, V^{(\omega)}_n) - z \big)^{-2\vartheta} \Big\|_1.
      \end{align}
      This estimate is just 
      the triangle inequality for the trace norm.
   \end{nummer}
\end{remarks}
The proof of Proposition~\ref{Prop:pore1} uses trace-ideal and resolvent techniques.
It is based on two lemmas.
The first one gives estimates on the Schatten $p$-norm of 
a function of the free Schr{\"o}dinger operator $ H_{\Lambda, \rm X}(0,0) $ 
times a multiplication operator
and on the trace of a power of the free resolvent. 
\begin{lemma}\label{Lemmafg}
  Let $ \Lambda \subset \mathbbm{R}^d $ be a bounded open cube and let ~$ \rm X = \rm D $ or
  $ \rm X =\rm N $.
  \begin{indentnummer*}
  \item 
    Let $ p \in [2, \infty[$, $  Q \in {\rm L}^\infty(\Lambda) $ and $ f: \mathbbm{R} \to \mathbbm{C} $ 
    be Borel measurable. Moreover, assume 
    $ f(H_{\Lambda, \rm X}(0,0)) \in  \mathcal{J}_p\big({\rm L}^2(\Lambda)\big) $.
    Then 
    $  f(H_{\Lambda, \rm X}(0,0)) \,  Q \in \mathcal{J}_p\big({\rm L}^2(\Lambda)\big) $ and 
    \begin{equation}\label{eq:fg}
      \big\| f(H_{\Lambda, \rm X}(0,0)) \,  Q \big\|_p 
      \leq 
      \left( \frac{2^{d}}{\left| \Lambda \right|}\right)^{1/p} \, 
      \big\|  f(H_{\Lambda, \rm X}(0,0)) \big\|_p 
      \,\, \big| Q \big|_{p}. 
      \end{equation}   
  \item 
    Let ~$ \alpha \in ]\, d/2, \infty[ $ and assume $ \left| \Lambda \right| \geq 1 $. Then 
     \begin{align}
         r(E_1,\alpha) 
         & := \frac{2^d}{\left| \Lambda \right|} \, 
         \big\| \left( H_{\Lambda, {\rm X}}(0, 0) - E_1 \right)^{-\alpha} \big\|_1 \\
         & \leq 
                \left| E_1 \right|^{\frac{d}{2} - \alpha} \frac{2^d}{(\alpha-1)!}
                \, \int_0^\infty \!\! \d \xi \, \e^{-\xi} \, \xi^{\alpha-1} \left( 1 + (2
                  \pi \xi)^{-1/2} \right)^d < \infty \notag
     \end{align} 
     for all ~$ E_1 \in ] -\infty , -1 ]$. 
   \end{indentnummer*}
\end{lemma}
\begin{remark}
  The validity of (\ref{eq:fg}) for all $ Q \in {\rm L}^\infty(\Lambda) $ may be extended to
  all $ Q \in {\rm L}^p(\Lambda) $ by an approximation argument.
  In this regard, Lemma~\ref{Lemmafg}(i) is a finite-volume analogue of \cite[Thm.~4.1]{Sim79TR}.
  However,  the bound in \cite[Thm.~4.1]{Sim79TR} is sharper than the one obtained by simply 
  taking the limit $ \Lambda \uparrow \mathbbm{R}^d $ in (\ref{eq:fg}). In fact, 
  in contrast to the version of the latter theorem for $ p = 2 $,
  equality can never hold in (\ref{eq:fg}).
  Nevertheless, the constant $ \left( 2^d/ \left| \Lambda \right| \right)^{1/p} $ in (\ref{eq:fg})
  is the best possible for all $ p \geq 2 $.
\end{remark}
\begin{proof}[ \bf Proof of Lemma~\ref{Lemmafg}]
  \begin{nummer} 
  \item 
    Let $ (\sgn g)(x) := g(x)/ | g(x) | $ 
    if $ g (x) \neq 0 $ and zero otherwise stand for the signum function of a complex-valued function $ g $.
    The polar decompositions $ Q = \left| Q \right| \sgn Q $  and 
    $ f(H_{\Lambda, \rm X}(0,0)) = \big| f(H_{\Lambda, \rm X}(0,0)) \big| \sgn f(H_{\Lambda, \rm X}(0,0)) $
    together with H{\"o}lder's inequality \cite[Eq.~(2.5b)]{Sim79TR} and \cite[Cor.~8.2]{Sim79TR} show that
    \begin{align}\label{eq:nochmalCor82}
      \big\| f(H_{\Lambda, \rm X}(0,0)) \,  Q \big\|_p^p 
      & \leq 
      \big\| \, \big|f(H_{\Lambda, \rm X}(0,0))\big| \,  \left| Q \right| \big\|_p^p
      \leq  
      \big\| \, \big|f(H_{\Lambda, \rm X}(0,0))\big|^{\frac{p}{2}} \,  
                                \left| Q \right|^{\frac{p}{2}} \big\|_2^2   \notag \\
      & = \Tr \big[ \left| Q \right|^{\frac{p}{2}} \big|f(H_{\Lambda, \rm X}(0,0))\big|^p  \left| Q \right|^{\frac{p}{2}} 
                          \big].
    \end{align}
    Let $ \{ \varphi_j \}_{j \in \mathbbm{N}} \subset  {\rm L}^2(\Lambda) $ denote an orthonormal eigenbasis
    associated with $ H_{\Lambda, \rm X }(0,0) $ and $ \varepsilon_j $ 
    the eigenvalue corresponding to $ \varphi_j  $.
    Then the trace in (\ref{eq:nochmalCor82}) may be calculated in this eigenbasis and estimated as follows
    \begin{equation}\label{eq:fgUgl}
      \sum_{j=1}^\infty \, \left| f(\varepsilon_j) \right|^p \,  
      \int_{\Lambda} \! \d^d x \, \left| \varphi_j(x) \right|^2 \, 
      \left| Q (x) \right|^p 
      \leq \frac{2^d}{\left| \Lambda \right|} \,   \big\| f(H_{\Lambda, \rm X}(0,0))\big\|_p^p 
      \,\,  \big| Q \big|_{p}^p.
      \quad
    \end{equation}
    The inequality is a consequence of the uniform boundedness 
    $ \left| \varphi_j \right|^2 \leq 2^d / \left| \Lambda \right| $ 
    for all $ j \in \mathbbm{N} $ which follows from the explicitly known expressions for $\{ \varphi_j \}$,
    see \cite[p.~266]{ReSi78}.
  \item  Using the integral represention of powers of resolvents,
    we get 
    \begin{equation}
       \big\| \left( H_{\Lambda, {\rm X}}(0, 0) - E_1 \right)^{-\alpha} \big\|_1 
       = \frac{1}{(\alpha-1)!} \,
      \int_0^\infty \! \d t \,\,t^{\alpha-1}\, \e^{t E_1}\,  \Tr\Big[ \e^{-t H_{\Lambda, \rm X}(0,0)}\Big].
    \end{equation}
    The claimed bound hence follows from the estimates  
     $\Tr\left[ \e^{ - t \, H_{\Lambda,{\rm X}}(0,0)} \right] 
     \leq
    \Tr\left[ \e^{ - t \, H_{\Lambda,{\rm N}}(0,0)} \right]$\hspace{0pt} 
    $ \leq |\Lambda| \big( |\Lambda|^{-1/d} + (2\pi t)^{-1/2}\big)^d  $ which are obtained by 
    Dirichlet-Neumann bracketing \cite[Prop.~4(b) on p.~270]{ReSi78} and the 
    explicitly known \cite[p.~266]{ReSi78} spectrum of $ H_{\Lambda,{\rm N}}(0,0) $.
  \end{nummer}
\end{proof}
The second lemma estimates Schatten norms of certain products involving bounded multiplication operators
and the resolvent of 
the magnetic Schr{\"o}dinger operator $ H_{\Lambda, \rm X}(a,0) $ without scalar potential.
\begin{lemma}\label{LemmaHS}
  Assume the situation of Proposition~\ref{Prop:pore1} and introduce
  \begin{equation}\label{eq:DefRa}
     R_{a} := \left( H_{\Lambda, {\rm X}}(a,0) - E_1 \right)^{-1} , \qquad  E_1 \in ] - \infty, -1].
  \end{equation}
  Then the following two assertions hold:
  \begin{nummer}
  \item  \label{LemmaHS1}
    Let $ p \in [2 , \infty[ $ and $ \alpha > 0 $ such that  $ \alpha p > d/2 $.
    Moreover, let  $ Q \in {\rm L}^{\infty}(\Lambda) $. Then 
    \begin{equation}\label{eq:resolventq}
      \big\| R_{a}^{\alpha} 
      %\left( H_{\Lambda, {\rm X}}(a,0) - E \right)^{-\alpha} 
      \, Q \big\|_p^p 
      \leq   r(E_1,\alpha p)  \,\, \big| Q \big|_{p}^p.
    \end{equation}
  \item \label{LemmaHS2}
    Let $ Q_1, \dots , Q_{k+1} \in {\rm L}^{\infty}(\Lambda) $. 
    Then
    \begin{equation} \label{Eq:Trace1}
      \big\| | Q_{k+1}|^{\frac{1}{2}} R_a\,  Q_k \cdots  R_a \, Q_1 \big\|_2^2 
      \leq  r(E_1,2k) \, \big| Q_{k+1} \big|_{2k+1} 
      \, \, \prod_{j=1}^k  \, \left|  Q_{j} \right|_{2k+1}^2. 
     \end{equation}
   \end{nummer}
\end{lemma}
\begin{proof}
  \begin{nummer}
  \item 
    The claim follows from the chain of inequalities
    \begin{align}
       \big\|  R_a^\alpha \, Q \big\|_p^p  
       \leq  \big\|  R_a^\alpha \, \left| Q \right| \big\|_p^p 
       & \leq  \big\|  R_a^{\frac{\alpha p}{2}} 
       \,  | Q |^{\frac{p}{2}} \big\|_2^2 \notag \\
       & \leq \big\|  R_0^{\frac{\alpha p}{2}} \,  | Q |^{\frac{p}{2}} \big\|_2^2
       \leq \frac{2^d}{\left| \Lambda \right|} \, \big\| R_0^{\alpha p} \big\|_1 \,
         \big| Q \big|_p^p.
    \end{align}
    Here the first inequality is a consequence of the 
    polar decomposition $ Q = \left| Q \right| \sgn{Q} $. The second one is a 
    special case of \cite[Cor.~8.2]{Sim79TR}. For the third one we used the  
    diamagnetic inequality \cite[Eq.~(A.23)]{HuLeMuWa01} in the version
     \begin{equation} 
       \left|   R_a^{\frac{\alpha p}{2}} 
       \,  | Q |^{\frac{p}{2}} \, \varphi \right| 
      \leq    R_0^{\frac{\alpha p}{2}} 
       \,  | Q |^{\frac{p}{2}} \,  \left|  \varphi \right| 
    \end{equation}
    for any $ \varphi \in {\rm L}^2(\Lambda) $, together with \cite[Thm.~2.13]{Sim79TR}. 
    The fourth inequality eventually follows from Lemma~\ref{Lemmafg}.
  \item 
    We repeatedly use H{\"o}lder's inequality \cite[Eq.~(2.5b)]{Sim79TR} for Schatten norms
    \begin{multline}
      \left\|  | Q_{k+1}|^{\frac{1}{2}} R_a \, Q_k \, \cdots  R_a  Q_1  \right\|_2
      \leq \prod_{j=1}^{k} \, 
      \Big\| \left| Q_{j+1} \right|^{\frac{j}{2k}} R_a \left| Q_{j} \right|^{\frac{2k+1-j}{2k}}\Big\|_{2k}
      \\
      \leq 
      \prod_{j=1}^{k} \, \Big\| R_a^{\frac{j}{2k+1}} \left| Q_{j+1} \right|^{\frac{j}{2k}} \Big\|_{\frac{2k(2k+1)}{j}}
         \, \, \Big\| R_a^{\frac{2k+1-j}{2k+1}} \left| Q_{j} \right|^{\frac{2k+1-j}{2k}}  \Big\|_{\frac{2k(2k+1)}{2k+1-j}}.
         \qquad
    \end{multline}
    The proof is completed using part~(i)
    of the present lemma. 
  \end{nummer}
\end{proof}
We are now ready to present a
\begin{proof}[ \bf Proof of Proposition~\ref{Prop:pore1}]
  The proof is split into the following three parts:
  \begin{list}{\labelnummer}{\usecounter{numcount}%
                  \topsep1ex\partopsep2ex\parsep0pt\itemsep1.5ex%
                  \labelwidth3em\itemindent0em\labelsep1em%
                  \leftmargin4em}%
   
    \item[{\bf a)}] Proof of part~(i) for $ v \in {\rm L}^{\infty}_{\rm loc}(\mathbbm{R}^d) $,
    \item[{\bf b)}] Proof of part~(ii) for $ v , v' \in {\rm L}^{\infty}_{\rm loc}(\mathbbm{R}^d) $,
    \item[{\bf c)}] Approximation argument for the validity
      of part~(i) and  (ii).
  \end{list}%\end{description}
  Throughout the proof 
  we use the abbreviations
  \begin{equation}
    R_{a,v}(z) := (H_{\Lambda, {\rm X}}(a, v) - z)^{-1} \quad \mbox{and} \quad  
    R_a = R_{a,0}(E_1),
  \end{equation}
  in agreement with (\ref{eq:DefRa}). \\
  
  \noindent
  {\bf As to a).} \hspace{0.1cm}
  Let $ v \in {\rm L}^{\infty}_{\rm loc}(\mathbbm{R}^d) $. We may then apply the (second)
  resolvent equation \cite{Weid80}
  \begin{equation}\label{Eq:resolvent}
    R_{a,v}(z)  =  R_a 
    + R_a \, Q \, R_{a,v}(z), \qquad Q:=  z - E_1 - v. 
  \end{equation}
  By the triangle inequality for the Schatten $p$-norm $ \| \cdot \|_p $, H{\"o}lder's inequality 
  and the standard estimate 
  $\left\| R_{a,v}(z) \right\| \leq \left| \Im z \right|^{-1} $, involving the usual (uniform) 
  operator norm 
  $ \| \cdot \| $, the resolvent equation 
  yields
  \begin{equation}
    \left\| R_{a,v}(z) \right\|_{p}
    \leq  \left\| R_a \right\|_{p} +  \left| \Im z \right|^{-1}  \left\| R_a  Q \right\|_{p}. 
  \end{equation}
  Using Lemma~\ref{LemmaHS1} we thus have
  \begin{align}
     \big\| \big( R_{a,v}(z) \big)^p \big\|_{1} & =  \big\| R_{a,v}(z) \big\|_{p}^p  \leq r(E_1,p) \, 
     \left( \left| \Lambda \right|^{1/p} +  \left| \Im z \right|^{-1} \,  \left| Q \right|_p \right)^p 
     \notag \\
     & \leq  r(E_1,p)  \left( 1  +  \left| \Im z \right|^{-1} \right)^p \, \big| 1 + | Q | \big|_p^p,
  \end{align}
  because $ \max\big\{ | \Lambda |^{1/d}, | Q |_p \big\} \leq \big| 1 + | Q | \big|_p $.
  The proof is finished by the upper bound in Lemma~\ref{Lemmafg}(ii).\\ 
  
  \noindent
  {\bf As to b).} \hspace{0.1cm}
  We start from the resolvent equation for powers of resolvents
  \begin{equation}\label{eq:itres}
    \big( R_{a,v}(z) \big)^{2k}
    - \big( R_{a,v'}(z) \big)^{2k} 
    = \sum_{j=1}^{2k} \, \big( R_{a,v'}(z) \big)^{2k+1-j}
    \left(v' - v\right) \big( R_{a,v}(z) \big)^{j},
  \end{equation}
  see \cite[Eq.~(5.4)]{PaFi92}. 
  Moreover, by the standard iteration of the resolvent equation~(\ref{Eq:resolvent}) 
  we have
  \begin{equation}\label{eq:Stoerungsreihe}
    \big( R_{a,v}(z) \big)^{j} = \sum_{r=0}^j \, \sum_{\substack{J \subseteq \{1, \dots , j\} \\ \# J = \, r}}
    R_a M_1(J) \cdots   R_a M_j(J) \, \big( R_{a,v}(z) \big)^{r}, 
  \end{equation}
  where $ M_s(J) := 1 $ if $ s \in J $ and $ M_s(J) := Q $ if $ s \not\in J $, and the second
  sum extends over all subsets $ J  \subseteq \{1, \dots , j\} $ with $ \# J = r $ elements.
  Using (\ref{eq:Stoerungsreihe}) and a suitable analogue with 
  $ v $ replaced by  $v'$ in (\ref{eq:itres}), 
  the trace norm of the l.h.s.\ of (\ref{eq:itres}) is seen to be bounded from above by a sum of
  finitely many terms of the form
  \begin{multline}\label{Eq:Rest2}
      \big\|  \,  \big( R_{a,v'}(z) \big)^s \, Q_1 R_a \cdots \, 
      Q_k\, R_a \, Q_{k+1}\, R_a \, Q_{k+2} 
      \cdots R_a Q_{2k+1} \, \big( R_{a,v}(z) \big)^r \big\|_1 \\
      \leq  \left| \Im z \right|^{-s-r} \, 
       \big\| Q_{1} \, R_a \cdots  Q_k\, R_a \, \left| Q_{k+1} \right|^{\frac{1}{2}}  \big\|_2
       \\
       \times \,
       \big\| \left| Q_{k+1} \right|^{\frac{1}{2}}\,  R_a  \, Q_{k+2} \cdots   R_a \, Q_{2k+1} \big\|_2,
       \quad
  \end{multline}
  with $ s $, $r \in \{ 0 , 1, \dots, 2k \} $.
  Each of these terms involves multiplication operators $Q_1, \dots , Q_{2k+1}$ suitably chosen from the
  set $ \{ 1 , z - E_1 - v , z - E_1 - v' , v' - v \} $ where \emph{exactly one}
  is equal to $ v' - v $. 
  The estimate (\ref{Eq:Rest2}) is again H{\"o}lder's inequality.
  The proof is then finished with the
  help of Lemma~\ref{LemmaHS2} and Lemma~\ref{Lemmafg}(ii), because $ v' - v $ appears
  only once and the other three operators in the above set 
  are all bounded by $ 1 + |z- E_1| + |v| $ since $ |v'| \leq |v| $.\\
   
  \noindent
  {\bf As to c).} \hspace{0.1cm}
  We approximate $ v \in {\rm L}^{2k+1}_{\rm loc}(\mathbbm{R}^d) $ by $ v_n^m $ defined through 
  \begin{equation}
    v_n^m(x) := \max\left\{ -n , \min\{ m , v(x) \}\right\}
  \end{equation}
  with $ x \in \mathbbm{R}^d $ and  $n$, $ m \in \mathbbm{N} $. Consequently, 
  we let $ v_\infty^m(x) := \min\{ m , v(x) \} $. Monotone (decreasing) convergence for forms
  \cite[Thm.~S.16]{ReSi80} yields the strong convergence
  \begin{equation}\label{eq:strongRes1}
    R_{a, v_\infty^m}(z) =  \mathop{\mbox{s-lim}}_{n \to \infty} \,  R_{a, v_n^m}(z)
  \end{equation}
  for all $ m \in \mathbbm{N} $ and all $ z \in \mathbbm{C} \backslash \mathbbm{R} $. 
  On the other hand, monotone (increasing) 
  convergence for forms \cite[Thm.~S.14]{ReSi80} yields 
   \begin{equation}\label{eq:strongRes2}
    R_{a, v}(z) =  \mathop{\mbox{s-lim}}_{m \to \infty} \,  R_{a, v_\infty^m}(z)
  \end{equation}
  for all $ z \in \mathbbm{C} \backslash \mathbbm{R} $. 
  We therefore have
  \begin{equation}\label{eq:ProbSim1}
    \big\| R_{a,v}(z) \big\|_p 
    \leq \limsup_{m \to \infty} \, \limsup_{n \to \infty}\,   \big\|  R_{a,v_n^m}(z) \big\|_p
  \end{equation} 
  where we used the non-commutative version of Fatou's lemma \cite[Prob.~167 on p.~385]{ReSi78} 
  (see also \cite[Thm.~2.7.(d)]{Sim79TR}) twice.
  Similarly, 
  \begin{equation}\label{eq:ProbSim2}
    \big\| \big( R_{a,v}(z) \big)^{2k} -  \big( R_{a,v'}(z) \big)^{2k} \big\|_1 
    \leq 
     \limsup_{m \to \infty} \, \limsup_{n \to \infty}\,  
     \big\| \big( R_{a,v_n^m}(z) \big)^{2k} -  \big( R_{a,{v'}_n^m}(z) \big)^{2k} \big\|_1
     \quad
  \end{equation}
  by the strong resolvent convergences (\ref{eq:strongRes1}) and (\ref{eq:strongRes2}), its
  analogue with $ v $ replaced by $ v' $ and \cite[Prob.~167 on p.~385]{ReSi78}.
  Applying part~a) and b) of the present proof to the pre-limit expressions in (\ref{eq:ProbSim1})
  and (\ref{eq:ProbSim2}) completes the proof of Proposition~\ref{Eq1:Prop1} and  \ref{Eq2:Prop1}. 
\end{proof}

%%%%%%%%%%%%%%%%%%%%%%%%%%%%%%%%%%%%%%%%%%%%%%%%%%%%%%%%%%%%%%%%%%%%%%%%%%%%%%%%%%%%%%%%%%%%%%%%
%%%%%%%%%%%%%%%%%%%%%%% PROPOSITION INFINITE VOLUME %%%%%%%%%%%%%%%%%%%%%%%%%%%%%%%%%%%%%%%%%%%%
%%%%%%%%%%%%%%%%%%%%%%%%%%%%%%%%%%%%%%%%%%%%%%%%%%%%%%%%%%%%%%%%%%%%%%%%%%%%%%%%%%%%%%%%%%%%%%%%
\subsection{Infinite-volume trace-ideal estimates}\label{Subsec:Aux2} 
It remains to prove the substitute of Lemma~5.12 (resp. 5.14) in
\cite{PaFi92}. It is the infinite-volume analogue of
Proposition~\ref{Prop:pore1} above. Accordingly, we will use the notation~(\ref{eq:notation}) 
with $ \Lambda = \mathbbm{R}^d $.
\begin{proposition}\label{Prop:pore2}
  Assume the situation of Theorem~\ref{Thm:IDOS} and recall the definition (\ref{Def:Vn}) of the
  truncated random potential $ V_n $.
  Let $ E_2 \in ] - \infty , 0[ $.
  Then for
  every $z \in \mathbbm{C}\backslash\mathbbm{R} $ there exist two
  constants $C_3(\Im z)$, $C_4(\Im z) > 0$, which depend on $ d $ and $ p $, but are independent of
  $\Gamma$, $n$, $A$, $V$, and $E_2$,
  such that
  \begin{nummer}
    \item\label{Eq:pore21}
      ~~$ \displaystyle
      \mathbbm{E}\Big\{ \big\| \indfkt{\Gamma} \left|  H(A,V_n) - z \right|^{-2\vartheta} 
      \indfkt{\Gamma} \big\|_1 \Big\} $\\[2ex]
      \hspace*{1.5cm}
      $ \displaystyle
      \leq C_3(\Im z) \,  | \Gamma | \,\, \left| E_2 \right|^{\frac{d}{2} - 2 \vartheta} \,
      \mathbbm{E}\Big[
      \int_{\Lambda(0)} \!\!\! \d^d x \,\, \left(
        1 + |z- E_2| + |V(x)| \right)^{2\vartheta} \Big]$.\\[2ex]
   \hspace*{1cm} The same holds true if $  V_n  $ is replaced by $ V $.
  \item\label{Eq:pore22}
     ~~$ \displaystyle
     \mathbbm{E}\Big\{ \big\| \indfkt{\Gamma} \big[   
     \left| H(A,V) - z \right|^{-2\vartheta} - \left|  H(A,V_n) - z \right|^{-2\vartheta} \big]
     \, \indfkt{\Gamma} \big\|_1 \Big\}$\\[2ex]
     \hspace*{1.5cm}
     $ \displaystyle
     \leq  C_4(\Im z) \,  | \Gamma | \,\, \left| E_2 \right|^{\frac{d}{2} - 2 \vartheta} \, 
     \Bigl\{ \mathbbm{E}\Big[ \int_{\Lambda(0)} \!\!\! \d^d x \,\, 
     ( 1  + |z - E_2| + |V(x)| )^{2\vartheta+1} \Big] \Bigr\}^{\frac{2\vartheta}{2\vartheta+1}}$\\[2ex]
     \hspace*{3cm}
     $ \displaystyle
     \times
     \,  \Bigl\{ \mathbbm{E}\Big[ \int_{\Lambda(0)} \!\!\! \d^d x \, |
     V(x) - V_n(x) |^{2\vartheta+1} \Big] \Bigr\}^{\frac{1}{2\vartheta+1}}$.
   \end{nummer}
\end{proposition}
\begin{remark}
  We recall from~(\ref{Def:nutilde}) that the l.h.s.\ of Proposition~\ref{Eq:pore21} coincides with 
  $ \widetilde  \nu_n(z, 2\vartheta) \, | \Gamma | $. Moreover, by the triangle inequality we have
  \begin{equation}\label{eq:infinitevol}
    \left| \, \widetilde \nu(z, 2\vartheta) - \widetilde \nu_n(z , 2\vartheta) \right| 
    \leq \frac{1}{| \Gamma | } \,   \mathbbm{E}\Big\{ \big\| \, \indfkt{\Gamma} \big[   
          \left| H(A,V) - z \right|^{-2\vartheta} - \left|  H(A,V_n) - z \right|^{-2\vartheta} \big]
        \, \indfkt{\Gamma} \big\|_1 \Big\}.
  \end{equation}
\end{remark}
The proof of Proposition~\ref{Prop:pore2} is split into two parts. In
the first part, the assertion is proven for $ \mathbbm{R}^d $-ergodic
random potentials. We will thereby closely follow \cite[Lemma~5.12/5.14]{PaFi92}.  
The $ \mathbbm{Z}^d $-ergodic case is treated
afterwards with the help of the so-called
\emph{suspension construction} \cite{Kir85b,Kir85a}.\\
\begin{proof}[ \bf Proof of Proposition~\ref{Prop:pore2} in case $ V $ is $ \mathbbm{R}^d $-ergodic]
  Throughout, we assume that $ V $ is $ \mathbbm{R}^d $-ergodic. The proof is split into three parts.
  \begin{list}{\labelnummer}{\usecounter{numcount}%
                  \topsep1ex\partopsep2ex\parsep0pt\itemsep1.5ex%
                  \labelwidth3em\itemindent0em\labelsep1em%
                  \leftmargin4em}%
    \item[{\bf a)}] Proof of part~(i),
    \item[{\bf b)}] Proof of part~(ii) with $ V $ replaced by $ V_m $ with $m \in \mathbbm{N}$ arbitrary, 
    \item[{\bf c)}] Approximation argument for the validity
      of part~(i) with $ V_n $ replaced by $ V $ and of part~(ii).
  \end{list}
  We use the abbreviations
  \begin{equation}
    R_{A,V}(z) := (H(A, V) - z)^{-1} \quad \mbox{and} \quad  
    R_A :=  R_{A,0}(E_2)
  \end{equation}
  for the resolvents of $ H(A, V) $ and $ H(A,0) $.\\

  \noindent
  {\bf As to a).} \hspace{0.1cm}
    We write $\big| R_{A,V_n}(z) \big|^{2 \vartheta} =  
      \big( R_{A,V_n}(\overline z)\big)^\vartheta  \big( R_{A,V_n}(z) \big)^\vartheta  $,
    where $\overline z$ is the complex conjugate of $ z \in \mathbb{C} $. 
    Suitably iterating the
    (second)  resolvent equation \cite{Weid80}
    \begin{equation}\label{Eq:resolvent2}
      R_{A,V_n}(z) = R_{A} + R_{A} \, Q \, R_{A,V_n}(z), \qquad Q:= z - E_2 - V_n,
    \end{equation} 
    we obtain the analogue of (\ref{eq:Stoerungsreihe}) for $ \big( R_{A,V_n}(z) \big)^\vartheta $.
    Using this equation and its adjoint, we are confronted with estimating finitely many
    terms of the form
    \begin{multline}\label{Eq:RestB2}
       \mathbbm{E} \Big\{ \big\| \indfkt{\Gamma} \, R^{(s)} \,
            \widetilde Q_1 \, R_A \cdots  \widetilde Q_\vartheta \, R_A \,
            R_A \, Q_\vartheta  \cdots  R_A \, Q_1 R^{(r)} \, 
            \indfkt{\Gamma} \big\|_1 \Big\} \\
       \leq  \,
       \Bigr\{  \mathbbm{E} \Big[ \big\| \indfkt{\Gamma} \, R^{(s)} \,
            \widetilde Q_1 \, R_A \cdots   \widetilde Q_\vartheta \, R_A \big\|_2^2 
                               \Big] 
            \, \,
            \mathbbm{E} \Big[ \big\|  
                                 R_A \, Q_\vartheta  \cdots  R_A \, Q_1 
                                 \, R^{(r)} \indfkt{\Gamma}  \big\|_2^2
                        \Big]
        \Bigl\}^{\frac{1}{2}}. \quad
     \end{multline}
     Here $s$, $r \in \{0, 1, \dots, \vartheta \} $ and 
     $ R^{(s)}$ denotes some product of $s$ factors each of which either being $ R_{A,V_n}(z)$ or its adjoint.
     Moreover, 
     $ \widetilde Q_1, \dots ,\widetilde Q_\vartheta $ respectively 
     $ Q_1, \dots , Q_\vartheta $ are random potentials suitably chosen from the set
     $ \{ 1, z - E_2 - V_n, \overline z - E_2 - V_n \}$. The estimate in (\ref{Eq:RestB2})
     is just H{\"o}lder's inequality for the
     trace norm and for the expectation. Thanks to \cite[Lemma~5.10]{PaFi92} we may use the estimate
     $ \big\| R^{(r)} \big\| \leq \left| \Im z \right|^{-r} $ inside the expectation. We therefore obtain the 
     inequality
     \begin{multline}\label{Eq:ResolvEx}
        \mathbbm{E} \Big[ \big\|  
                                 R_A \, Q_\vartheta  \cdots  R_A \, Q_1 
                                 \, R^{(r)} \indfkt{\Gamma} \big\|_2^2
                        \Big] \leq \left| \Im z \right|^{-2r} \,
                        \mathbbm{E} \Big[ \big\|  
                                 R_A \, Q_\vartheta  \cdots  R_A \, Q_1 \indfkt{\Gamma} 
                                 \big\|_2^2  \Big] \\
        \leq   \left| \Im z \right|^{-2r} \,
                        \mathbbm{E} \Big[ \big\|  
                                 R_0 \, \left| Q_\vartheta \right|  
                                 \cdots  R_0 \, \left| Q_1 \right| \indfkt{\Gamma} 
                                  \big\|_2^2  \Big],
                                  \qquad
     \end{multline}
     and analogously for the other factor, involving $  R^{(s)} $ instead of $  R^{(r)} $.
     The second inequality in (\ref{Eq:ResolvEx}) is a consequence of \cite[Thm.~2.13]{Sim79TR}
     and the 
     diamagnetic inequality \cite{Sim79,HuLeMuWa01} which upon iteration gives 
     \begin{equation}
       \big| R_A \, Q_\vartheta  \cdots  R_A \, Q_1 \, \indfkt{\Gamma} \varphi \big| 
       \leq  R_0 \, \left| Q_\vartheta \right|  
              \cdots R_0 \, \left| Q_1 \right| \, \indfkt{\Gamma}  \left| \varphi \right|
     \end{equation}
     for all $ \varphi \in {\rm L}^2(\mathbbm{R}^d)$. To complete the proof we use the iterated H{\"o}lder  
     inequality as in \cite[Lemma~5.11(i)]{PaFi92}. 
     Taking there $p=2\vartheta$, $g_j = g_{p+2-j} = \left| Q_j \right| $, $ t_j = t_{p+2-j} = 2 \vartheta $
     for $ j \in \{1, \dots , \vartheta \} $ and $g_p =1$, $ t_p = \infty $, we in fact obtain
     \begin{equation}\label{eq:Lemma5.11a}
        \mathbbm{E} \Big[ \big\|  
                                 R_0 \, \left| Q_\vartheta \right|  
                                 \cdots R_0 \, \left| Q_1 \right| \indfkt{\Gamma} 
                                  \big\|_2^2  \Big] 
        \leq C_5(E_2) \, \left| \Gamma \right| \, \prod_{j=1}^{\vartheta} \, \Big\{ 
                        \mathbbm{E}\Big[ \left| Q_j(0) \right|^{2\vartheta} \Big] \Big\}^{\frac{1}{\vartheta}},
     \end{equation}
     with
     \begin{equation}
       C_5(E_2) :=  \int_{\mathbbm{R}^d} \frac{\d^d p}{(2 \pi )^{d}} \, 
       \Big(\frac{p^2}{2} - E_2 \Big)^{-2 \vartheta} 
       = \left| E_2 \right|^{d/2 - 2 \vartheta} \, 
       \frac{(2 \vartheta -1 - \frac{d}{2})!}{(2 \pi )^{\frac{d}{2}} (2\vartheta -1)!},
     \end{equation}
     see also \cite[Lemma~5.9]{PaFi92}.
     Since $ \max_{j}\{ | Q_j | , | \widetilde Q_j | \}\leq 1 + |z-E_2| + |V| $ 
     for all $ j \in \{1, \dots , \vartheta \} $ 
     the proof is complete.\\
  
  \noindent
  {\bf As to b).} \hspace{0.1cm}   
  We let $ m$, $ n \in \mathbbm{N}$. The resolvent equation for powers of resolvents gives
  \begin{align}
    & \big| R_{A,V_m}(z) \big|^{2\vartheta} -  \big| R_{A,V_n}(z) \big|^{2\vartheta}  \notag \\
    & \qquad =  \sum_{k=1}^{2\vartheta}\,  \left(\, \prod_{i=1}^k R_{A,V_m}(z_i)
    \right) \left(V_n - V_m\right) \left(\, \prod_{j=k}^{2\vartheta} R_{A,V_n}(z_j) \right), \qquad
  \end{align}  
  see also \cite[Eq.~(5.4)]{PaFi92}, with $z_k = z$ if $k \in \{ 1 , \dots , \vartheta \} $
  and $z_k =\overline z$ otherwise. Using the resolvent equation (\ref{Eq:resolvent2}) and 
  its adjoint, we may accumulate in total $ 2 \vartheta $ resolvents $ R_A $, analogously 
  to what was done to obtain (\ref{Eq:RestB2}), such that
  we are confronted with estimating finitely many terms of the form
  \begin{multline}\label{eq:part2hoelder}
     \mathbbm{E} \Big\{ \big\| \indfkt{\Gamma} \, R^{(s)} \, Q_1 \, R_A \cdots 
     Q_{\vartheta} \, R_A \, Q_{\vartheta+1} \,  R_A \, Q_{\vartheta+2}  
     \cdots  R_A \, Q_{2\vartheta+1}  R^{(r)} \,  \indfkt{\Gamma}
                        \big\|_1 \Big\} \\
     \leq \, \Bigr\{  \mathbbm{E} \Big[ \big\| \indfkt{\Gamma}  \, R^{(s)} \, Q_1 \, R_A \cdots
     Q_{\vartheta} \, R_A \, \left| Q_{\vartheta+1} \right|^{\frac{1}{2}} \big\|_2^2 
                                \Big] \qquad \\
                                \times
                                \,
                   \mathbbm{E} \Big[ \big\| \left| Q_{\vartheta+1} \right|^{\frac{1}{2}} 
                       R_A \, Q_{\vartheta+2}  \cdots  R_A \, Q_{2\vartheta +1} R^{(r)}
                            \indfkt{\Gamma} \big\|_2^2 \Big] 
           \Bigl\}^{\frac{1}{2}}. \quad
  \end{multline}
  Here $ s $, $ r \in \{ 0, 1, \dots,  2 \vartheta \} $ and
  $ R^{(s)} $ is some product of $ s $ factors each of which either being 
  $ R_{A,V_m}(z) $, $ R_{A,V_n}(z) $ or one of their adjoints.
  Moreover, $ Q_1, \dots , Q_{2 \vartheta + 1} $ are random potentials
  suitably chosen from
  the set $ \{ 1 , z - E_2 - V_n , \overline z - E_2 - V_n,
   z - E_2 - V_m , \overline z - E_2 - V_m, V_n - V_m \}$ 
  and exactly one of these is equal to $ V_n - V_m $.
  We now copy the steps between (\ref{Eq:ResolvEx}) and (\ref{eq:Lemma5.11a}) and take
  $p=2\vartheta$, $g_j = g_{p+2-j} = \left| Q_j \right| $,  $g_p =  \left| Q_{\vartheta +1} \right|$ and 
  $ t_j = 2 \vartheta +1$
  for $ j \in \{1, \dots , \vartheta +1 \} $ in \cite[Lemma~5.11(i)]{PaFi92} to obtain the bound
  \begin{multline}\label{eq:Lemma5.11b}
        \mathbbm{E} \Big[ \big\|  
         \left| Q_{\vartheta+1} \right|^{\frac{1}{2}} \, R_0 \, \left| Q_\vartheta \right|  
                                 \cdots  R_0 \, \left| Q_1 \right| \indfkt{\Gamma} 
                                  \big\|_2^2  \Big] \\ 
        \leq C_5(E_2) \, \left| \Gamma \right| 
         \,
        \Big\{ \mathbbm{E}\Big[ 
                 \left| Q_{\vartheta+1}(0) \right|^{2\vartheta+1} \Big] 
                 \Big\}^{\frac{1}{2\vartheta+1}}
        \, \prod_{j=1}^{\vartheta} \, 
        \Big\{ 
                \mathbbm{E}\Big[ \left| Q_j(0) \right|^{2\vartheta+1} \Big] 
                \Big\}^{\frac{2}{2\vartheta+1}}
        \quad
   \end{multline}
   for $\left| \Im z \right|^s $ times the first expectation on the r.h.s.\ of (\ref{eq:part2hoelder}). 
   The second expectation is treated similarly.
   Since exactly one of the $Q_j$ is equal to $ V_n - V_m $  
   and all others in the above set may be bounded
   by $ 1 + |z- E_2| + |V| $, the proof is complete.\\

  \noindent
  {\bf As to c).} \hspace{0.1cm} 
  Since $ H(A, V_n^{(\omega)})\, \varphi  \to  H(A, V^{(\omega)})\, \varphi $ as $ n \to \infty $
  for all $ \varphi \in \mathcal{C}_0^\infty(\mathbbm{R}^d) $ and
  $ \mathcal{C}_0^\infty(\mathbbm{R}^d) $ 
  is a common core for all $ H(A, V_n^{(\omega)}) $ and $  H(A, V^{(\omega)}) $ for
  $ \mathbbm{P} $-almost all $ \omega \in \Omega $ by Proposition~\ref{Prop:DefH1}, 
  \cite[Thm.~VIII.25(a)]{ReSi80} implies that 
  \begin{equation}\label{eq:stronresconver}
    R_{A,V^{(\omega)}}(z) =  \mathop{\mbox{s-lim}}_{n \to \infty} \,   R_{A,V_n^{(\omega)}}(z)
  \end{equation}
  for all $ z \in \mathbbm{C} \backslash  \mathbbm{R} $ and $ \mathbbm{P} $-almost all $ \omega \in \Omega $.
  Part (ii) of the present proof together with assumption~\ass{I} shows that
  \begin{equation}
    \lim_{n,m \to \infty} \, \int_{\Omega} \!\!  \mathbbm{P}(\d\omega) \,\, \Big\| \indfkt{\Gamma} 
                 \Big[ \big| R_{A,V_n^{(\omega)}}(z)\big|^{2\vartheta}   -  \big| R_{A,V_m^{(\omega)}}(z)\big|^{2\vartheta}
                 \Big] \indfkt{\Gamma} \Big\|_1 = 0.
  \end{equation}
  Analogous reasoning as in the proof of Corollary~\ref{Cor:fFormel} yields the existence of 
  some sequence $( n_{j} )$ of natural numbers such that
  \begin{equation}
    \lim_{i,j \to \infty} \, \Big\| \indfkt{\Gamma} 
                 \Big[ \big| R_{A,V_{n_i}^{(\omega)}}(z)\big|^{2\vartheta}   
                 -  \big| R_{A,V_{n_j}^{(\omega)}}(z)\big|^{2\vartheta}
                 \Big] \indfkt{\Gamma} \Big\|_1 = 0
  \end{equation}
  for $\mathbbm{P}$-almost all $\omega\in \Omega$.
  In other words, the subsequence 
  $ \big(  \indfkt{\Gamma} \big| R_{A,V_{n_j}^{(\omega)}}(z)\big|^{2\vartheta}  \indfkt{\Gamma} \big)_{j\in\mathbbm{N}} $ 
  is Cauchy in $ \mathcal{J}_1({\rm L}^2(\mathbbm{R}^d)) $ for $ \mathbbm{P}$-almost all $ \omega \in \Omega $.
  Thanks to completeness of $ \mathcal{J}_1({\rm L}^2(\mathbbm{R}^d)) $ and 
  the strong convergence (\ref{eq:stronresconver}), 
  we have the convergence
  \begin{equation}\label{eq:ConvTeilJ}
    \lim_{j \to \infty} \, \Big\| \indfkt{\Gamma} 
                 \Big[ \big| R_{A,V_{n_j}^{(\omega)}}(z)\big|^{2\vartheta}   -  \big| R_{A,V^{(\omega)}}(z)\big|^{2\vartheta}
                 \Big] \indfkt{\Gamma} \Big\|_1 = 0
  \end{equation}
  in $  \mathcal{J}_1({\rm L}^2(\mathbbm{R}^d)) $ for $\mathbbm{P}$-almost all $ \omega \in \Omega $.
  The latter implies
  \begin{equation}
    \int_{\Omega} \!\! \mathbbm{P}(\d\omega) \,\,  
    \Big\| \indfkt{\Gamma} \big| R_{A,V^{(\omega)}}(z)\big|^{2\vartheta} \indfkt{\Gamma} \Big\|_1 
    \leq \liminf_{j \to \infty}
    \int_{\Omega} \!\!  \mathbbm{P}(\d\omega) \,\, 
    \Big\| \indfkt{\Gamma} \big| R_{A,V_{n_j}^{(\omega)}}(z)\big|^{2\vartheta}  \indfkt{\Gamma} \Big\|_1
  \end{equation}
  by Fatou's lemma. Since Proposition~\ref{Eq:pore21} holds for all $ V_{n_j} $, the 
  proof of Proposition~\ref{Eq:pore21} with $ V_n $ replaced by $ V $ is complete.
  For a proof of Proposition~\ref{Eq:pore22} we proceed analogously using (\ref{eq:ConvTeilJ}), 
  part~(ii) of the present proof
  and again Fatou's lemma.
\end{proof}
It remains to carry over the result for $ \mathbbm{R}^d $-ergodic
potentials to $ \mathbbm{Z}^d $-ergodic ones using the suspension
construction, see \cite{Kir85b,Kir85a}.
\begin{proof}[ \bf Proof of Proposition~\ref{Prop:pore2} in case $ V $ is $ \mathbbm{Z}^d $-ergodic]
  We consider the product of the probability spaces $ ( \Omega ,
  \mathcal{A}, \mathbbm{P} ) $ and $ (\Lambda(0),
  \mathcal{B}(\Lambda(0)), \mathit{Lebesgue} )$. The latter corresponds to a
  uniform distribution on the open unit cube $ \Lambda(0) $.  On this
  enlarged space we define the random potential
\begin{equation}\label{Def:Vsusp}
        \mathcal{V} : \;
                       \left( \Omega \times \Lambda(0) \right) \times
                        \mathbbm{R}^d   \to  \mathbbm{R},  \qquad
                        \left( \omega , y , x \right)   \mapsto 
                        \mathcal{V}^{(\omega, y)}(x) := V^{(\omega)}(x - y).
\end{equation}
It is $\mathbbm{R}^d$-ergodic by construction
\cite{Kir85b} and enjoys properties~\ass{S} and \ass{I}.  The
latter assertion is proven by tracing the claimed properties of
$\mathcal{V}$ back to the respective properties of $V$.

It remains to prove that the validity of Proposition~\ref{Prop:pore2} 
for $ \mathcal{V} $ implies the one for $ V $.  For
this purpose, we note that the integral transform~(\ref{Def:nutilde})
of the (infinite-volume) density-of-states measure corresponding to $
\mathcal{V} $ obeys
\begin{align}\label{Eq:Gerechne}
  \frac{1}{\left|\Gamma \right|} \, & \int_{\Omega \times \Lambda(0)}
  \! \! \! \! \! \! \! \!  \mathbbm{P}(\d\omega) \otimes \d^d y \,\,
  \Tr \left[ \indfkt{\Gamma} \, \big| H(A, \mathcal{V}^{(\omega,y)}) -
    z \big|^{-2\vartheta}
    \, \indfkt{\Gamma} \right]  \notag \\
  = & \, \frac{1}{\left|\Gamma \right|} \, \int_{\Lambda(0)}\! \!
  \d^d y \, \int_{\Omega} \!\!  \mathbbm{P}(\d\omega) \, \Tr \left[
    \indfkt{\Gamma-y} \, \big| H(A, V^{(\omega)}) - z \big|^{-2\vartheta}
    \, \indfkt{\Gamma-y} \right]  \notag \\
  = & \, \frac{1}{\left|\Gamma \right|} \, \int_{\Omega} \!\! 
  \mathbbm{P}(\d\omega) \, \Tr \left[ \indfkt{\Gamma} \, \big| H(A,
    V^{(\omega)}) - z \big|^{-2\vartheta} \, \indfkt{\Gamma} \right]
\end{align}
Here the first equality results from (\ref{eq:magntrans}) and the definitions of $ \mathcal{V} $
and the cube $ \Gamma-y := \{ x - y \in \mathbbm{R}^d \; : \, x \in \Gamma \}$ together with Fubini's
theorem.  To obtain the second equality we have used the fact that the
trace does not depend on $ y $ after performing the 
$ \mathbbm{P}(\d \omega)$-integration. This follows from $ \mathbbm{Z}^d $-homogeneity as well as from the fact 
that one may 
``re-arrange'' $ \Gamma-y $ in the form of
$ \Gamma $ by $ \mathbbm{Z}^d $-translations since $ \Gamma $ is compatible with the lattice.  Moreover,
one computes
\begin{multline}
  \int_{\Omega \times \Lambda(0)} \! \! \! \! \! \! \! \! 
  \mathbbm{P}(\d\omega) \otimes \d^d y \,\,\left(
    1 + |z- E_2| + |\mathcal{V}^{(\omega,y)}(0)| \right)^{2\vartheta} \\
  = \int_{\Omega } \!\!  \mathbbm{P}(\d\omega) \, \int_{\Lambda(0)}
  \!\!\! \d^d x \,\, \left( 1 + |z- E_2| + |V^{(\omega)}(x)|
  \right)^{2\vartheta}
\end{multline}
using (\ref{Def:Vsusp}) and Fubini's theorem. This completes the proof
of Proposition~\ref{Eq:pore21} with $ V_n $ replaced by $ V $.  
The other parts of Proposition~\ref{Prop:pore2} in 
the $ \mathbbm{Z}^d $-ergodic case are proven
similarly.  
\end{proof}
%-------------------------------------------------------------------------
\section*{Acknowledgment}\addcontentsline{toc}{section}{Acknowledgments}
It's a pleasure to thank Kurt Broderix ($1962-2000$), Eckhard Giere, J{\"o}rn Lembcke, and
Georgi D.\ Raikov 
for helpful remarks and stimulating discussions.
The present work was supported by the Deutsche Forschungsgemeinschaft under
grant no. Le 330/12 which is a project within the
Schwerpunktprogramm ``Interagierende stochastische Systeme von hoher
Komplexit{\"a}t'' (DFG Priority Programme SPP 1033).
%------------------------------------------------------------------------------

%
%\newpage
%\citationindex
%------------------------------------------------------------------------------
\end{document}